\documentclass[a4paper,11pt]{article}
\pdfoutput=1
\usepackage{jheppub}
\usepackage{array}
\usepackage{color}
\newcommand\ii{\'{\i}}

\def\Dsl{\,\raise.15ex\hbox{/}\mkern-13.5mu D}

\newcommand{\beq}{\begin{eqnarray}}
\newcommand{\eeq}{\end{eqnarray}}

\newcommand{\nn}{\nonumber}

\newcommand{\te}{$S_{\rm top}$\ }

\newcommand{\eref}[1]{(\ref{#1})}

\title{Topological entropy and renormalization group flow in 3-dimensional spherical spaces}
\author[a]{M. Asorey,}
\author[b]{C. G. Beneventano,}
\author[a c]{I. Cavero-Pel\'aez}
\author[b]{D. D'Ascanio}
\author[b]{and \\E. M. Santangelo}
\affiliation[a]{Departamento de F\ii sica Te\'orica, Universidad de Zaragoza,\\
E-50009  Zaragoza, Spain }
\affiliation[b]{Departamento de F\ii sica, Universidad Nacional de La Plata,\\
Instituto de F\ii sica de La Plata, CONICET-Universidad Nacional de La Plata,\\
C.C.67, 1900 La Plata, Argentina}
\affiliation[c]{CUD,\\ E-50090, Zaragoza, Spain}

\emailAdd{asorey@unizar.es}
\emailAdd{gabriela@fisica.unlp.edu.ar}
\emailAdd{cavero@unizar.es}
\emailAdd{dascanio@fisica.unlp.edu.ar}
\emailAdd{mariel@fisica.unlp.edu.ar}
\abstract{
We analyze the renormalization group (RG) flow of the temperature independent term of the entropy
in the high temperature limit $\beta/a\ll1$
of a massive field theory in 3-dimensional  spherical  spaces, $M_3$, with constant curvature $6/a^2$.
For masses  lower  than $\frac{2\pi}{\beta}$, this term can be identified with the free energy
of the same theory on  $M_3$ considered as a  3-dimensional  Euclidean space-time.
The non-extensive part of this free energy, $S_{\rm hol}$, is generated by the holonomy of
the spatial metric connection. We show that for  homogeneous spherical spaces the
holonomy entropy $S_{\rm hol}$ decreases monotonically when the RG scale flows to the
infrared.  At the conformal fixed points  the values of the holonomy entropy do coincide with the
genuine topological entropies recently introduced. The monotonic behavior of the RG flow leads to an inequality  between the topological entropies of the conformal field theories connected by such flow, i.e. $S_{\rm top}^{UV}>S_{\rm top}^{IR}$. From a 3-dimensional viewpoint the same term arises in the 3-dimensional Euclidean effective action and has the same monotonic behavior under the RG group flow.
We conjecture that such monotonic  behavior is generic,
which would give rise to a 3-dimensional generalization of the c-theorem, along the lines of
the  2-dimensional  $c$-theorem  and  the 4-dimensional $a$-theorem.
The conjecture is related to  recent formulations of the $F$-theorem.
In particular,  the holonomy entropy on lens spaces is directly related to the
topological R\'enyi  entanglement entropy on  disks of 2-dimensional flat spaces.
}

\begin{document}
\maketitle
\flushbottom

\section{Introduction}

The RG flow in the space of quantum {fi}eld theories exhibits a number of interesting dynamical features. Fixed points of the RG flow correspond to scale invariant theories which usually are also conformally invariant \cite{pol1,pol2}. In the even-dimensional case, due to the existence of conformal anomalies (see \cite{wanomaly94} for a quite comprehensive review), in gravitational backgrounds conformal field theories induce an effective gravitational action which uniquely characterizes the  conformal background theory. In 1+1 dimensions  the gravitational effective action is uniquely characterized by the central charge $c$, which is the numerical factor of the Einstein term in the anomaly.
In four dimensions the anomaly has three independent parameters $(c,a, b)$ which multiply the Weyl tensor squared term $W^2$, the Euler density term $E$ and a surface term $\square R$, respectively. The first two coefficients $c$ and $a$ are universal and characterize the conformal theory. The last coefficient $b$ is renormalization dependent and is not universal \cite{shapiro}, which can be understood by the fact that it can be generated by a local term $R^2$ in the effective gravitational action that has to be determined from renormalization prescriptions.

Conformal invariance is usually broken by interactions. Wilson's renormalization group (RG) flow shows how the different theories evolve under scale transformations towards conformal theories, which are invariant fixed points of the flow. Since the RG flow involves a change of scale, to some extent ultraviolet (UV) singularities are smoothed, and the effects of massive degrees of freedom disappear from the effective theories as one approaches the infrared (IR) limit. Such  behavior is intrinsically linked to the irreversible nature of the RG flow and can be encoded by a monotonic entropy function.
This fact is the essence of the celebrated Zamolodchikov $c$-theorem \cite{zama}, which asserts that in two-dimensional field theories there exists a monotonic $c$-function in the space of field theories that is always decreasing as the RG flow points towards the IR.  The value of this function at conformal fixed points does coincide with the central charge $c$ of their conformal algebra, implying the inequality $c_{UV}>c_{IR}$ between the central charges of conformal field theories connected by RG flow trajectories.

After many attempts at generalizing the $c$-theorem to even-dimensional theories \cite{cardy, osborn, osborn-jack}, a consistent proof for the four-dimensional case has emerged very recently \cite{KS,K}. In this case, there exists a function with the suited behavior under the RG flow that, at the conformal fixed points, coincides with the coefficient $a$ of the conformal anomaly. Also in four dimensions, examples were encountered where the other
anomaly coefficient, $c$, does not have the desired behavior \cite{Cappelli:1990yc,Anselmi:1997am}.

Entropic theorems of this kind hold only in even space-time dimensions due to the even nature of trace anomalies.
In odd dimensions, there is no trace anomaly and the same approach does not work.
There have been attempts to derive other entropic theorems based on gravitational Chern-Simons densities generated by parity anomaly (see e.g. \cite{kogan}) without much success.
Recently, similar entropic functions were found  in the context of the AdS/CFT correspondence, and it has been conjectured that such a c-theorem might hold in odd-dimensional theories \cite{sm1,sm2}.
In particular, for three-dimensional Euclidean conformal field theories the finite, volume independent part, $F$, of the effective action of the theory conformally mapped to $S^3$ \cite{klebanov} has been proposed as an entropy function ($F$-theorem). In this picture the relevant quantity $F$ is related to a similar constant term which appears in the entanglement entropy of the field theory when tracing out degrees of freedom on a subdomain of the physical space \cite{kitaev}. This constant term is  universal \cite{ryu} and decreases along the trajectories
of the RG flow. By using this connection, a proof of the $F$-theorem was found in \cite{cas} (see also \cite{Liu:2012eea}).

In a previous paper \cite{Asorey:2012vp} we found a similar universal quantity associated to a subleading term
in the high temperature expansion of the entropy of some conformal invariant field theories on Euclidean space-times of the form $S^1\times M_3$.
This entropy is reminiscent of the boundary entropy which arises in one-dimensional manifolds with boundary \cite{aj2,cardy1,affleck,affleck2}. The boundary entropy, unlike the bulk entropy, follows an area law and mimics the black-hole entropy behavior, which means that in  1+1 dimensions it is dimensionless and can have a logarithmic dependence on the space volume or be a volume-independent quantity.  In a manifold without boundary the entropy has no logarithmic dependence on the volume, but dimensionless subleading contributions can arise.

In higher odd-dimensional space-times similar features occur. In three dimensions the entropy exhibits the same behavior at high temperature. Besides the extensive terms proportional to the volume of the space $M_3$, which include the usual Stefan-Boltzmann term, there is an extra, temperature-independent term which contains the topological entropy.
The full temperature-independent term can be split into two parts: one which depends on the spatial volume or can be expressed as an integral of a density, and another one which is independent of the spatial volume and essentially depends on the topology of $M_3$. It is this latter term that defines the {\it topological entropy} \cite{Asorey:2012vp}.

The topological entropy \te is a good candidate for a c-theorem in 3 dimensions. In this paper we analyze the behavior of \te under the RG flow for some theories on homogeneous spherical spaces, $M_3 = S^3/\Gamma^\ast$. These spaces are obtained as quotient spaces of the 3-dimensional sphere $S^3$ by finite subgroups $\Gamma^\ast$ of its fixed-point-free isometries.
We consider the mass perturbation of a conformal scalar field theory and analyze the dependence on the mass of the different thermodynamic quantities. In particular, we focus on the mass dependence of the non-extensive, temperature-independent term of the entropy $S_{\rm hol}(m)$ which only depends on the holonomy group of the Levi-Civita connection on $M_3$. In the limit of infinite (IR) or zero (UV) mass this term gives rise to the topological entropy of the corresponding conformal field theory. We analyze the monotonic behavior of this quantity $S_{\rm hol}(m)$ under the RG flow and show that it follows a pattern similar to that of any physical entropy, like the leading extensive terms of the bulk entropy.

\section{Massive scalar fields on spherical backgrounds}

Let us consider a scalar field $\phi$ with mass $m$ and arbitrary coupling $\xi$ to the curvature of the spherical space $S^3/\Gamma^\ast$.
In four space-time dimensions, the action functional for such a field is
\beq
S = \frac12 \int d^4x \sqrt{g} \left\{ g^{\mu\nu}\partial_{\mu}\phi\, \partial_{\nu}\phi + \frac{R}{6}\phi^2 + \left[ m^2 + \left(\xi - \frac{1}{6} \right) R \,\right] \phi^2\right\},
\eeq
where $g^{\mu\nu}$ is the $\mathbb{R}\times S^3/\Gamma^\ast$ metric and $R=6/a^2$ the scalar curvature of $S^3/\Gamma^\ast$. The first two terms of the action are conformally invariant, whereas the last term explicitly breaks scale symmetry. From this Einstein static metric we can recover the standard Friedmann-Lema\^{\i}tre-Robertson-Walker cosmological metric via a conformal mapping. The thermodynamics of fields on static Einstein space-times was first studied by Dowker and Critchley \cite{DC} and, in de Sitter space-times, by Gibbons and Hawking \cite{gh}.

We shall focus on the conformal coupling case $\xi=\frac16$. Notice, however, that the results can be straightforwardly generalized to arbitrary couplings as far as $ m^2 + \left(\xi - \frac{1}{6} \right) R \geq 0$. The extension to other cases (like zero mass with minimal coupling, for example) is less direct.

A finite temperature $T=1/\beta$ can be introduced in the standard way by compactifying the Euclidean time into a circle $S^1$ of radius $\frac\beta{2 \pi}$. The partition function is given by the determinant
\beq
Z(\beta) = \det \left( -\partial_0^2 - \Delta_c + m^2 \right)^{-\frac12},
\eeq
where $\partial_0$ is the Euclidean time derivative operator with periodic boundary conditions in $[0,\beta]$ and $\Delta_c=\Delta-1/a^2$ is the conformal Laplacian on the spherical spatial manifold. The determinant can be computed via the zeta function regularization method \cite{DC}. In this framework, the   effective action is given by
\beq\label{seff}
S_{\mathrm{eff}}(\beta) = -\log Z(\beta) = - \frac12 \left. \frac{d}{ds} \zeta(s) \right\vert_{s=0}.
\eeq

The eigenvalues of the operator $-\partial_0^2 - \Delta_c + m^2$ are the sum of the temporal $S^1$ modes given by the Matsubara frequencies
\beq
\omega_l^2 = \left(\frac{2\pi l}{\beta}\right)^2, \quad l\in\mathbb{Z},
\eeq
and  the eigenvalues of (minus) the conformal Laplacian with a mass term,
\beq
\lambda_k = \left(\frac{k}{a}\right)^2+m^2, \quad k=1,2,\ldots,\infty.
\label{eigenvaluesb}
\eeq
The degeneracies of the spatial modes are:
\begin{itemize}
\item for the $S^3$ background, $d_k = k^2$;
\item for an even order lens space $L(2q,1)\ =S^3/\mathbb{Z}_{2q}$,
\begin{equation}
     d_k =\begin{cases} 0,&  \quad \mbox{for} \quad k \quad \mbox{even} \\
      k \left(2\left[\frac{k}{2q}\right]+1\right),& \quad \mbox{for} \quad  k \quad \mbox{odd;}\end{cases}
      \label{evenw}
\end{equation}
\item for an odd order lens space $L(2q+1,1)\ =S^3/\mathbb{Z}_{2q+1}$,
\beq\label{odd}
d_k= \begin{cases}\displaystyle  k \ \frac{k-r}{2q+1},&\mbox{for}\quad r\quad \mbox{even}\\[2mm]
\displaystyle k\left(1+\frac{k-r}{2q+1}\right),&\mbox{for}\quad r\quad \mbox{odd;}\end{cases}
\eeq
where  $k=(2q+1)\left[\frac{k}{2q+1}\right]+r$.
{ {
\item for a prism space $S^3/D_{4p}^*$,
\begin{equation}
d_{2n+1} =\begin{cases} (2n+1)\left[\frac{n}{p}\right] ,&  \quad \mbox{for} \quad n \quad \mbox{odd} \\[3mm]
      (2n+1)\left(\left[\frac{n}{p}\right]+1\right),& \quad \mbox{for} \quad  n \quad \mbox{even.}\end{cases}\\
      \label{prism}
\end{equation}
}}
\end{itemize}

The zeta function of the relevant operator then reads
\beq\label{zeta}
\zeta(s) = \mu^{2s} \sum_{\substack{k=1\\l=-\infty}}^{\infty} d_k \left[ \left(\frac{k}{a}\right)^2 + m^2 + \left(\frac{2\pi l}{\beta}\right)^2 \right]^{-s}.
\eeq
As usual, we have introduced a scale parameter $\mu$ to render the zeta function dimensionless. From geometrical considerations \cite{Blau:1988kv}, it can be shown that the zeta function does not vanish at $s=0$, so the $s$-derivative in \eqref{seff} does depend on $\mu$. We shall show later how this dependence can be removed by means of a natural physical renormalization prescription.

We shall focus mainly in the high temperature limit. However, since the theory has many couplings  ($m,a,\beta$), it is convenient to consider the different asymptotic regimes from a very general perspective.

In the next two sections we shall perform analytic extensions of the zeta function \eqref{zeta} for the case of the sphere in two different regimes: low ($\beta\rightarrow \infty$) and high ($\beta\rightarrow 0$) temperatures. In both regimes we shall distinguish between small mass and large mass regimes.

\section{Effective action on $S^1\times S^3$: low temperature regime $\frac{\beta}{a}\gg 1$}

In a spherical background the calculation is the simplest one, since the degeneracies of the spatial modes scale with a power law.
The zeta function \eref{zeta} for the relevant operator on the sphere,
\beq
\zeta_{S^3}(s) = \left( \frac{\mu\beta}{2\pi} \right)^{2s} \sum_{\substack{k=1\\l=-\infty}}^{\infty} k^2 \left[ \left(\frac{k\beta}{2\pi a}\right)^2 + \left(\frac{\beta m}{2\pi}\right)^2 + l^2 \right]^{-s},
\eeq
can be rewritten, by using Schwinger's proper-time representation, as
\beq
\zeta_{S^3}(s) = \left( \frac{\mu\beta}{2\pi} \right)^{2s} \frac{1}{\Gamma(s)} \sum_{\substack{k=1\\l=-\infty}}^{\infty} k^2 \int_0^{\infty}dt\, t^{s-1} e^{-\left[ \left(\frac{k\beta}{2\pi a}\right)^2 + \left(\frac{\beta m}{2\pi}\right)^2 +l^2 \right]t}.
\eeq

The Poisson summation formula (which, in this case, is the inversion formula for the Jacobi theta function), when applied to the $l$-sum, leads to
\beq
\zeta_{S^3}(s) = \left( \frac{\mu\beta}{2\pi} \right)^{2s} \frac{\pi^{\frac12}}{\Gamma(s)} \sum_{\substack{k=1\\l=-\infty}}^{\infty} k^2 \int_0^{\infty}dt\, t^{s-\frac12-1} e^{-\left[ \left(\frac{k\beta}{2\pi a}\right)^2 + \left(\frac{\beta m}{2\pi}\right)^2 \right]t - \frac{\left(\pi l\right)^2}{t}}.
\eeq
The integration of the terms $l=0$ and $l\neq 0$ yields \cite{rusa}
\beq
\zeta_{S^3}(s) &=& \left( \frac{\mu\beta}{2\pi} \right)^{2s} \frac{\pi^{\frac12}}{\Gamma(s)} \left\{ \Gamma(s-\tfrac12) \sum_{k=1}^{\infty} k^2 \left[ \left( \frac{k\beta}{2\pi a} \right)^2 + \left( \frac{\beta m}{2\pi} \right)^2 \right]^{\frac12-s}  \right.\\ \nn
& &  +  \left. 4\sum_{k,l=1}^{\infty} k^2 (\pi l)^{s-\frac12} \left[ \left( \frac{k\beta}{2\pi a} \right)^2 + \left( \frac{\beta m}{2\pi} \right)^2 \right]^{\frac14-\frac{s}{2}} \!\!\!\!K_{s-\frac12}\left( \sqrt{k^2 + (am)^2} \frac{\beta}{a} l\right)
\right\},
\eeq
where $K_{s-\frac12}(x)$ is the modified Bessel function of order ${s-\frac12}$.

\bigskip

The contribution of the Bessel function $K_{\frac12}(x)$ to the $l\neq 0$ terms of the effective action \eqref{seff} can be written as
\beq
S_{\mathrm{eff}, S^3}^{l\neq 0}(\beta) = \sum_{k=1}^{\infty} k^2 \log{\left( 1 - e^{-\frac{\beta}{a}\sqrt{k^2+(am)^2}} \right)}.
\label{seffbajasl}
\eeq

The $l=0$ term,
\beq
\zeta^{l=0}_{S^3}(s) = \left( \frac{\mu\beta}{2\pi} \right)^{2s} \frac{\pi^{\frac12}}{\Gamma(s)} \Gamma(s-\tfrac12) \sum_{k=1}^{\infty} k^2 \left[ \left( \frac{k\beta}{2\pi a} \right)^2 + \left( \frac{\beta m}{2\pi} \right)^2 \right]^{\frac12-s},
\label{zetabajas0}
\eeq
must be analytically extended in order to have a convergent sum  for the effective action. This can be achieved in two ways, each one being convenient for taking a different mass limit, i.e., $m\to0$ and $m\to\infty$ respectively.

\subsection{Large mass-volume case $am \gg 1$}

Let us analyze the $l=0$ contribution in \eqref{zetabajas0}. It can be rewritten as
\beq
\zeta^{l=0}_{S^3}(s) = (\mu a)^{2s} \frac{\beta}{4\pi^{\frac12}a} \frac{\Gamma(s-\tfrac12)}{\Gamma(s)} \sum_{k=-\infty}^{\infty} \left\{ \left[ k^2 + (am)^2 \right]^{\frac32-s} - (am)^2 \left[ k^2 + (am)^2 \right]^{\frac12-s} \right\}.
\eeq
Using Schwinger's proper-time representation for each term and the Poisson summation formula in the $k$-sum we obtain
\beq\nn
\zeta^{l=0}_{S^3}(s) = \frac{(\mu a)^{2s}}{\Gamma(s)}\frac{\beta}{4a} &&\left\{\frac12 (am)^{4-2s} \,\Gamma(s-2) \right. + 4\left(s-\tfrac32\right) \sum_{k=1}^{\infty} \left( \frac{\pi k}{am} \right)^{s-2} K_{s-2}(2\pi amk)  \\
& & - \left. 4(am)^2\sum_{k=1}^{\infty} \left(\frac{\pi k}{am}\right)^{s-1} K_{s-1}(2\pi amk) \right\},
\eeq
for $ma\neq 0$.

The  $s$-derivative of this contribution can be derived from the small-$s$ expansion of $\Gamma(s-2)/\Gamma(s)$,
\beq\label{gamma}
\frac{\Gamma(s-2)}{\Gamma(s)} = \frac{1}{2} + \frac{3s}{4}+ O(s^2).
\eeq
The result is
\beq\nn
S_{\mathrm{eff}, S^3}^{l=0}(\beta) &=& -\frac{(am)^4}{32}\frac{\beta}{a} \left[ \frac{3}{2} + 2\log\frac{\mu}{m} \right]
\\\label{seffbajas0}
& & + \frac{(am)^3}{2\pi} \frac{\beta}{a} \sum_{k=1}^{\infty} \frac{1}{k} K_1(2\pi amk) + \frac{3(am)^2}{4\pi^2} \frac{\beta}{a} \sum_{k=1}^{\infty} \frac{1}{k^2} K_2(2\pi amk).
\eeq

We remark that the  effective action does depend on the scale parameter $\mu$, as already argued. The problem  is now how to renormalize this quantity or, in other words, which terms to subtract in order to obtain a physically meaningful result. We note that the contributions to the effective action corresponding to the terms in the first line of equation \eqref{seffbajas0} do not vanish for any value of $\mu$ in the infinite volume limit ($a\rightarrow\infty$), where we should get a free theory in flat space-time at zero temperature. A natural  renormalization prescription is, then, to remove these terms from the effective action. Thus, we add to the bare action a counterterm of the form
\beq\label{counter}
\Delta S=\frac{(am)^4}{32}\frac{\beta}{a} \left[ \frac{3}{2} + 2\log\left(\frac{\mu}{m}\right) \right],
\eeq
in order to match the normal ordering prescription in a flat space-time background.

With this renormalization prescription the renormalized effective action becomes
\beq\nn
S_{\mathrm{eff}, S^3}(\beta) &=& \frac{3(am)^2}{4\pi^2}\frac{\beta}{a} \sum_{k=1}^{\infty} \frac{1}{k^2} K_2(2\pi amk) + \frac{(am)^3}{2\pi}\frac{\beta}{a} \sum_{k=1}^{\infty} \frac{1}{k} K_1(2\pi amk)  \\ \label{seff1}
& & + \sum_{k=1}^{\infty} k^2\log\left(1-e^{-\frac{\beta}{a}\sqrt{k^2+(am)^2}}\right).
\eeq

The expression of the vacuum (Casimir) energy derived from \eref{seff1} when
$\beta \rightarrow \infty$,
\beq
E_C &=& \frac{3(am)^2}{4\pi^2 a} \sum_{k=1}^{\infty} \frac{1}{k^2} K_2(2\pi amk) + \frac{(am)^3}{2\pi a} \sum_{k=1}^{\infty} \frac{1}{k} K_1(2\pi amk) \,,\label{ce}\eeq
is exponentially suppressed for large masses, showing that the decoupling theorem works in this case with the chosen renormalization prescription \cite{Bordag:1996ma}. For the same reason, the entropy is exponentially suppressed in this regime and there is no temperature-independent contribution.

Notice that expression \eref{ce} agrees with the one obtained, via the sum of modes, in \cite{elizalde}.

\subsection{Small mass-volume case $am < 1$}

For $am < 1$  we rewrite \eqref{zetabajas0} as
\beq
\zeta^{l=0}_{S^3}(s) = (\mu a)^{2s} \frac{\beta}{2\pi^{\frac12} a} \frac{\Gamma(s-\tfrac12)}{\Gamma(s)} \sum_{k=1}^{\infty} k^{3-2s} \left[ 1+ \left( \frac{am}{k} \right)^2 \right]^{\frac12 -s},
\eeq
and make use of the binomial expansion to obtain
\beq
\zeta^{l=0}_{S^3}(s) = (\mu a)^{2s} \frac{\beta}{2\pi^{\frac12} a} \frac{\Gamma(s-\tfrac12)}{\Gamma(s)} \sum_{n=0}^{\infty} \frac{\Gamma\left(\tfrac32 -s\right)}{n!\,\Gamma\left(\tfrac32 -n-s\right)} (am)^{2n} \zeta_R(2s+2n-3),
\eeq
where $\zeta_R(z)$ is the Riemann zeta function. The $s$-derivative can be computed from the small-$s$ expansion of $\zeta_R(2s+1)$,
\beq\label{riemann}
\zeta_R(2s+1) = \frac{1}{2s}+\gamma + O(s).
\eeq

The result is
\beq
S_{\mathrm{eff}, S^3}^{l=0}(\beta) = \frac{\beta}{2a} \sum_{\substack{n=0\\n\neq 2}}^{\infty} \frac{\Gamma(\tfrac32)}{n!\,\Gamma(\tfrac32-n)}(am)^{2n}\zeta_R(2n-3) - \frac{(am)^4}{16}\frac{\beta}{a} \left[ \log\left(\frac{\mu a}{2}\right){+1}+\gamma) \right].
\eeq

If we perform the same subtraction \eref{counter} as in the large volume-mass regime, the renormalized   effective action reads
\beq\label{seff2}
S_{\mathrm{eff}, S^3}(\beta) &=& \frac{\beta}{240a} - \frac{(am)^2}{48}\frac{\beta}{a} - \frac{(am)^4}{16}\frac{\beta}{a} \left[ \log\left(\frac{m a}{2}\right){+\frac14}+\gamma \right]  \\ \nn
& & + \frac{\beta}{2a} \sum_{n=3}^{\infty} \frac{\Gamma\left(\tfrac32\right)}{n!\,\Gamma\left(\tfrac32 -n\right)}(am)^{2n}\zeta_R(2n-3) + \sum_{k=1}^{\infty} k^2\log\left(1-e^{-\frac{\beta}{a}\sqrt{k^2+(am)^2}}\right).
\eeq
We remark that in the conformal $m=0$ limit the expression \eref{seff2} agrees with equation (4.2) of reference \cite{Asorey:2012vp}. In the same way the vacuum energy at zero temperature is recovered from the effective action \eref{seff2} and agrees with known results obtained by using the same renormalization scheme in the large mass regime (see appendix \ref{ap1} for a detailed discussion of the mode sum).
Notice that for any value of the mass  the entropy is exponentially suppressed and there is no temperature-independent contribution in the zero-temperature limit.

\section{Effective action on $S^1\times S^3$: high temperature regime $\frac{\beta}{a}\ll 1$}

We have analyzed above two different expressions for the effective action of the scalar theory on the sphere $S^3$ that are useful in the low temperature regime. We have found a renormalization prescription which gives a finite Casimir energy for any range of masses and  fulfills the requirement of  decoupling in the infinite mass limit at zero temperature. However, in order to study the behavior of the topological entropy under the RG flow we have to deal with the high temperature regime.
To this purpose, we first rewrite equation (\ref{zeta}) as
\beq\nn
\zeta_{S^3}(s) &=& \mu^{2s} a^2\sum_{\substack{k=1\\l=-\infty}}^{\infty} \left[ \left(\frac{k}{a}\right)^2 + m^2 + \left(\frac{2\pi l}{\beta}\right)^2 \right]^{-s+1} \\  & & -\mu^{2s} a^2 \sum_{\substack{k=1\\l=-\infty}}^{\infty} \left(m^2 + \left(\frac{2\pi l}{\beta}\right)^2 \right)\left[ \left(\frac{k}{a}\right)^2 + m^2 + \left(\frac{2\pi l}{\beta}\right)^2 \right]^{-s}.
\eeq
Using Schwinger's proper-time representation for each term and Poisson summation in the $k$-sum we obtain
\beq
\zeta_{S^3}(s) &=& \frac{ {\pi}^{\frac12}({\mu a})^{2s}}{2} \sum_{l=-\infty}^{\infty}\left\{\left[(am)^2+{\left(\frac{2\pi a l}{\beta}\right)}^2\right]^{-s+\frac32}\frac{\Gamma(s-\frac32)}{2\Gamma(s)}\right.\\ \nn
& & +\frac{4}{\Gamma(s-1)}\sum_{k=1}^{\infty} (k\pi)^{s-\frac32}\!\!\left[(am)^2+{\left(\frac{2\pi a l}{\beta}\right)}^2\right]^{-\frac{s}{2}+\frac34}\!\!\!
K_{s-\frac32}\!\!\left(2k\pi\sqrt{(am)^2+{\left(\frac{2\pi a l}{\beta}\right)}^2}\right)\\ \nn
& & - \left. \frac{4}{\Gamma(s)}\sum_{k=1}^{\infty} (k\pi)^{s-\frac12}\!\!\left[(am)^2+{\left(\frac{2\pi a l}{\beta}\right)}^2\right]^{-\frac{s}{2}+\frac54}\!\!\!
K_{s-\frac12}\!\!\left(2k\pi\sqrt{(am)^2+{\left(\frac{2\pi a l}{\beta}\right)}^2}\right)\right\}\,.
\eeq

Now, the $k=0$ contribution must be analyzed in a different way for the $m\beta< 2\pi$ and $m\beta> 2\pi$ regimes.

\subsection{Small mass/temperature case ${m\beta}<{2\pi}$}

In this case, separating the $l=0$ term in the $k=0$ contribution and performing a binomial expansion in the $l\neq 0$ sum, the zeta function can be rewritten as
\beq\nn
\zeta^{k=0}_{S^3}(s) &=& (\mu a)^{2s} \frac{\pi^{\frac12}}{4} \frac{\Gamma\left(s-\tfrac32\right)}{\Gamma(s)} \Bigg[ (am)^{3-2s}   \\
& &  \left. + 2 \left(\frac{2\pi a}{\beta}\right)^{3-2s} \sum_{n=0}^{\infty} \frac{\Gamma\left(\tfrac52 - s\right)}{n!\,\Gamma\left(\tfrac52-n-s\right)} \left(\frac{m\beta}{2\pi}\right)^{2n} \zeta_R(2s+2n-3) \right].
\eeq
The $s$-derivative can be obtained from the small-$s$ expansion \eqref{riemann}, resulting
\beq\nn
S_{\mathrm{eff}, S^3}^{k=0}(\beta) &=& -\frac{\pi}{6} (am)^3 -\frac{\pi}{3} \left(\frac{2\pi a}{\beta}\right)^3 \sum_{\substack{n=0\\n\neq 2}}^{\infty} \frac{\Gamma\left(\tfrac52\right)}{n!\,\Gamma\left(\tfrac52-n\right)}\left(\frac{m\beta}{2\pi}\right)^{2n}\zeta_R(2n-3) \\ & & -\frac{(am)^4}{32}\frac{\beta}{a} \left[ 2\log\left(\frac{\mu\beta}{2\pi}\right) - 3\psi(1) + \psi\left(\tfrac12\right) \right].
\eeq

The sum of  $k=0$ and $k\neq 0$ contributions and counterterms \eqref{counter} of the renormalization prescription yields the renormalized effective action
\beq\label{seff4}
S_{\mathrm{eff}, S^3}(\beta)\!\! &=&\!\! -\frac{\pi}{6} (am)^3 -\frac{\pi^4}{45}\left(\frac{a}{\beta}\right)^3\!\! + \frac{\pi^2}{12} \frac{a^3 m^2}{\beta} - \frac{(am)^4}{32}\frac{\beta}{a} \left[ 2\log\left(\frac{m\beta}{2\pi}\right) - 3\psi(1) + \psi\left(\tfrac12\right) -\frac32 \right]  \nn\\
& &\!\! -\frac{\pi}{3} \left(\frac{2\pi a}{\beta}\right)^3 \sum_{n=3}^{\infty} \frac{\Gamma\left(\tfrac52\right)}{n!\,\Gamma\left(\tfrac52-n\right)}\left(\frac{m\beta}{2\pi}\right)^{2n}\zeta_R(2n-3) +  \frac{1}{4\pi^2} \sum_{\substack{l=-\infty \\ k=1}}^{\infty} \frac{1}{k^3} \\ \nn & & \!\!\times  \left[ 1 + 4\pi^2 \sqrt{l^2 + \left(\tfrac{m\beta}{2\pi}\right)^2} \frac{a}{\beta} k + \frac12 \left(4\pi^2 \sqrt{l^2 + \left(\tfrac{m\beta}{2\pi}\right)^2} \frac{a}{\beta} k\right)^2\right] e^{-4\pi^2 \sqrt{l^2 + \left(\tfrac{m\beta}{2\pi}\right)^2} \frac{a}{\beta} k}.
\eeq
The leading terms of the high temperature expansion are
\beq\label{seff4high}
S_{\mathrm{eff}, S^3} &=& -\frac{\pi}{6} (am)^3 -\frac{\pi^4}{45}\left(\frac{a}{\beta}\right)^3 + \frac{\pi^2}{12} (am)^2 \frac{a}{\beta}\nonumber\\
&+& \frac{1}{4\pi^2}\sum_{k=1}^{\infty} \frac{1}{k^3} \left[ 1 + 2\pi ma k + \frac12 (2\pi mak)^2 \right] e^{-2\pi ma k}+{\cal O}({\textstyle \frac\beta{m}}).\eeq

In the $m=0$ conformal limit, this result coincides with the one obtained in \cite{Asorey:2012vp}. In particular, the value of the temperature-independent terms of \eref{seff4high}
\beq\label{seff4highb}
 -\frac{\pi}{6} (am)^3
&+& \frac{1}{4\pi^2}\sum_{k=1}^{\infty} \frac{1}{k^3} \left[ 1 + 2\pi ma k + \frac12 (2\pi mak)^2 \right] e^{-2\pi ma k}\eeq
agrees, when $m=0$, with the value $\zeta(3)/4 \pi^2$ obtained in \cite{Asorey:2012vp}. However, in this massive case with $\frac{m\beta}{2\pi} <1$, besides the usual Planck term ${\cal O}(\frac{a^3}{\beta^3})$, there is an extra subleading contribution ${\cal O}(\frac{a}{\beta})$ to the entropy, as well as new temperature-independent terms that contribute to the topological entropy (and which were missing in \cite{Elizalde:2003cv}).
These subleading, temperature-independent terms of \eref{seff4highb} only depend on the remaining parameters of the theory, i.e., the mass $m$ and the radius of the sphere $a$. As in the conformal $m=0$ case \cite{Asorey:2012vp,Dowker12}, they coincide with the corresponding terms in the effective action of the same field theory defined on the lower dimensional Euclidean space-time $S^3$ \cite{dhs,Dowker14}.

We shall see in  next section that the infinite temperature limit is entirely different when $m$ grows faster than $2\pi/\beta$.

\subsection{Large mass/temperature case ${m\beta}\geq{2\pi}$}

Let us finally analyze the case where, even if the temperature is high, the mass of the field is larger than the temperature. In that case, instead of using a binomial expansion in the $k=0$ contribution, we   perform a
Poisson-Jacobi inversion in the Matsubara modes sum, $l$, which yields
\beq\nn
\zeta^{k=0}_{S^3} (s) = \frac{(\mu a)^{2s}}{\Gamma(s)} \frac{\pi}{4} \!\!\left( \frac{2\pi a}{\beta} \right)^{\!\!3-2s} \!\!\left[ \left(\frac{m\beta}{2\pi}\right)^{4-2s} \!\!\Gamma(s-2) + 4\sum_{l=1}^{\infty} \left(\frac{2\pi^2l}{m\beta}\right)^{s-2} \!\!K_{s-2}\left(m\beta l\right) \right],
\eeq
 for $m\beta\neq 0$. Using the expansion  \eqref{gamma} once more, it is straightforward to obtain the corresponding contribution to the effective action,
\beq
S_{\mathrm{eff}, S^3}^{k=0}(\beta) = -\frac{(am)^4}{32} \frac{\beta}{a} \left[\frac32 + 2\log\left(\frac{\mu}{m}\right)\right] - (am)^2\frac{a}{\beta} \sum_{l=1}^{\infty} \frac{1}{l^2} K_2\left(m\beta l\right).
\eeq

This expression has to be renormalized within the same scheme used throughout the paper, i.e., by adding the counterterm \eref{counter}. The result is
\beq\label{seff3}
S_{\mathrm{eff}, S^3}(\beta) &=& -(am)^2 \frac{a}{\beta} \sum_{l=1}^{\infty} \frac{1}{l^2} K_2\left(m\beta l\right) +  \frac{1}{4\pi^2} \sum_{\substack{l=-\infty \\ k=1}}^{\infty} \frac{1}{k^3} \\ \nn
& & \times \left[ 1 + 4\pi^2 \sqrt{l^2 + \left(\tfrac{m\beta}{2\pi}\right)^2} \frac{a}{\beta} k + \frac12 \left(4\pi^2 \sqrt{l^2 + \left(\tfrac{m\beta}{2\pi}\right)^2} \frac{a}{\beta} k\right)^2\right] e^{-4\pi^2 \sqrt{l^2 + \left(\tfrac{m\beta}{2\pi}\right)^2} \frac{a}{\beta} k}.
\eeq

This last expression shows that, for $m\beta\rightarrow \infty$ when $\beta \rightarrow 0$ (i.e. $m=\mathcal{O}\left(\frac{1}{\beta^{1 + \epsilon}}\right),\, \epsilon>0$), the effective action vanishes in the infinite temperature limit. Thus, particles with a mass greater than any other dimensional parameter in the theory decouple not only at zero temperature but also in the high temperature limit. It is possible to obtain a different limit by fine-tuning a dependence of $m$ on $\beta$ such that $m\beta$ remains finite as $\beta \rightarrow 0$. In such case, one has
\beq
S_{\mathrm{eff}, S^3}(\beta) &\sim & -(m\beta)^2 \left(\frac{a}{\beta}\right)^3 \sum_{l=1}^{\infty} \frac{1}{l^2} K_2\left(m\beta l\right)\, ,\eeq
where we have disregarded exponentially suppressed terms. This last expression holds not only for $m\beta \geq 2\pi$ but also for $m\beta < 2\pi$. In the latter case it does also agree with \eref{seff4}, and it provides its analytic extension beyond the domain of convergence (${m\beta}<{2\pi}$) of the series in the fifth term.

Notice that in the infinite mass limit with $m\gg\frac{1}{\beta}\gg \frac{1}{a}$ the temperature-independent terms vanish (as does the whole effective action), which does not agree with the effective action of the same field theory defined on the lower dimensional Euclidean space-time $S^3$, unlike the case of ${m\beta}<{2\pi}$. In this last case, the temperature-independent terms do coincide with the effective action on $S^3$ even when $m$ goes to infinity, provided that it grows slower than $\frac{2\pi}{\beta}$. The difference between these two regimes cannot be understood in terms of the pure 3-dimensional theory, but only from a four-dimensional approach where the temperature $1/\beta$ introduces a scale which allows to discriminate between the two regimes.

\section{Effective action on spherical spaces}\label{lens}

In order to analyze the behavior of the topological entropy under the RG flow we shall consider scalar fields on
homogeneous spherical spaces. The classification of compact Riemannian spaces with positive constant curvature is
very well known (see e.g. Wolf \cite{wolf}) and it is given in terms of the quotients of the 3-dimensional sphere $S^3$ by the discrete groups of fixed-point-free isometries which are in one-to-one correspondence with the discrete subgroups of $SU(2)$ via the identification $S^{3}\equiv SU(2)$. That is to say, $M_3=S^3/\Gamma^\ast$, where $\Gamma^\ast$ can be either a cyclic group $\mathbb{Z}_p$ (which gives rise to a lens space $L(p,1)$), a binary dihedral group $D_{4p}^*$ of order $4p$ (giving rise to a prism space), and the binary tetrahedral $T^*$, octahedral $O^*$ or icosahedral $I^*$ groups, of order 24, 48 and 120 respectively (giving respectively the octahedral, truncated cube and Poincar\'e dodecahedral spaces).
We shall focus on the leading terms of the effective action at high temperature on those spherical spaces.

In the case of lens spaces, because of the differences in the spectrum of the Laplace-Beltrami
operator \cite{ikeda, Dowker041} it is convenient to distinguish between odd-$p$ and even-$p$ cases.

\subsection{Odd-order lens spaces $L(2q+1,1)$}

The zeta function \eqref{zeta} with the odd lens spaces degeneracies \eqref{odd} can be written as
\beq\nn
\zeta_{L(2q+1,1)}(s) &=& \mu^{2s}\sum_{\substack{n=0 \\ l=-\infty}}^{\infty} \left[  \sum_{\substack{r=0\\ r \mathrm{\,even}}}^{2q} (n(2q+1)+r) n + \sum_{\substack{r=1\\ r \mathrm{\,odd}}}^{2q-1} (n(2q+1)+r) \left(n+1\right)  \right]  \left( \lambda_k + \omega_l^2 \right)^{-s} \\[2mm]\label{zetaodd}
& = & \frac{1}{2q+1}\zeta_{S^3}(s)+\delta\zeta_{L(2q+1,1)}(s),
\eeq
where $k=n(2q+1)+r$ with $r=0,1,\dots, 2q$, and
\beq\nn
& & \delta\zeta_{L(2q+1,1)}(s) \!\!=\!\! -\frac{\mu^{2s}}{2q+1}\!\!\sum_{\substack{n=0 \\ l=-\infty}}^{\infty}\!\! \left\{ \sum_{r=1}^{q} 2r \!\left[n(2q+1)\!+\!2r\right] \!\left[\left(\frac{n(2q+1)+2r}{a}\right)^2 \!+\!m^2 \!+\!\left(\frac{2\pi l}{\beta}\right)^2 \right]^{-s}  \right. \\ \nn
& & \,\,\,\,\,\,\,\,\,\,\,\,\,\,\,+ \left.\sum_{r=0}^{q-1} 2(r-q) \left[n(2q+1)+2r+1\right] \left[\left(\frac{n(2q+1)+2r+1}{a}\right)^2 +m^2 +\left(\frac{2\pi l}{\beta}\right)^2 \right]^{-s}\right\}.
\eeq
Now, changing $q-r\rightarrow r$ in the second $r$-sum and extracting convenient overall factors one gets
\beq\nn
& & \delta\zeta_{L(2q+1,1)}(s)  = -2\left(\frac{\mu a}{2q+1}\right)^{2s}\sum_{\substack{n=0 \\ l=-\infty}}^{\infty} \sum_{r=1}^{q} r
\nn \\ & &\,\,\,\,\,\,\times \left\{ \left(n+\frac{2r}{2q+1}\right) \left[\left(n+\frac{2r}{2q+1}\right)^2 +\left(\frac{am}{2q+1}\right)^2 +\left(\frac{2\pi a l}{(2q+1)\beta}\right)^2 \right]^{-s} \right. \\ \nn
& & \,\,\,\,\,\,- \left. \left(n+1-\frac{2r}{2q+1}\right) \left[\left(n+1-\frac{2r}{2q+1}\right)^2 +\left(\frac{am}{2q+1}\right)^2 +\left(\frac{2\pi a l}{(2q+1)\beta}\right)^2 \right]^{-s}\right\}.
\eeq

We remark that, by inverting the sum over $l$ one can easily see that this part of the zeta function vanishes at zero temperature. Therefore, we do not have to change the counterterm in order to renormalize the effective action; it is the same as in the case of $S^1 \times S^3$ divided by the volume factor of the lens space $2q+1$. This is a generic property of all physical observables.  UV divergences can only arise in covariant local terms of the effective action, and due to spherical symmetry, the corresponding  local densities are constant and identical for  all spherical spaces with the same curvature.
Thus, in the same scheme of UV renormalization the absolute values of the counterterms
for the different spherical spaces only differ in a volume factor, which in the case of
lens spaces  $L(2q+1,1)$ is  given by the multiplicative factor $1/(2q+1)$ relative to the sphere $S^3$.

At high temperature, it is easy to see that the contribution of higher Matsubara modes ($l\neq 0$) to $\delta\zeta_{L(2q+1,1)}(s)$ leads to terms in the effective action that are exponentially suppressed as the temperature increases, $$\exp\left(-\frac{2\pi a n}{2q+1}\sqrt{m^2+\left(\frac{2\pi l}{\beta}\right)^2}\right).$$
The only contributions that are not exponentially suppressed are the trivial Matsubara modes $l=0$
\beq\nn
& &\delta\zeta_{L(2q+1,1)}^{l=0} (s)\! =\! { -2\left(\frac{\mu a}{2q\!+\!1}\right)^{2s}}\!\!\sum_{n=-\infty}^{\infty} \! \sum_{r=0}^{q-1}\! (r\!+\!1)\! \left(n\!+\!\frac{2r\!+\!2}{2q\!+\!1}\right) \left[\left(n\!+\!\frac{2r\!+\!2}{2q\!+\!1}\right)^2 \!+\!\left(\frac{am}{2q\!+\!1}\right)^2 \right]^{-s} \\ \nn
& & ={ -\frac{2q+1}{2(1-s)}\left(\frac{\mu a}{2q+1}\right)^{2s}}\!\!\! \sum_{n=-\infty}^{\infty}\sum_{r=0}^{q-1} {(r+1)} \left.\frac{d}{d\alpha}\right\vert_{\alpha=1} \left[\left(n+\frac{2r+2\alpha}{2q+1}\right)^2 +\left(\frac{am}{2q+1}\right)^2 \right]^{-s+1}\!\!\!\!\!\!\!\!\!.
\eeq
The $n$-sum can be performed via a Poisson inversion using the Schwinger proper-time representation in all terms. The result of the integrals gives
\beq\nn
& &\delta\zeta_{L(2q+1,1)}^{l=0} (s) \!=\!{ -\frac{2q+1}{2(1\!-\!s)\Gamma(s\!-\!1)}\!\left(\frac{\mu a}{2q\!+\!1}\right)^{2s}}\sum_{r=0}^{q-1}{\! (r\!+\!1)}\! \left.\frac{d}{d\alpha}\right\vert_{\alpha=1} \left[\pi^{\frac12}\Gamma(s\!-\!\tfrac32) \left(\frac{am}{2q\!+\!1}\right)^{3-2s}  \right. \\
 & & \left.{ + 4\pi^{s-1} \left(\frac{am}{2q+1}\right)^{\frac32-s}}\sum_{n=1}^{\infty} n^{s-\frac32}\cos{\left(2\pi\frac{2r+2\alpha}{2q+1}n\right)}K_{s-\frac32}\left(\frac{2\pi am n}{2q+1}\right)\right] \\ \nn
 &=&{ -\frac{8 s}{\Gamma(s+1)}\left[\frac{\pi\mu^2 a}{(2q+1)m}\right]^s \left(\frac{am}{2q+1}\right)^{\frac32} }\sum_{r=1}^{q}r \sum_{n=1}^{\infty} n^{s-\frac12}\sin{\left(\frac{4\pi rn}{2q+1}\right)}K_{s-\frac32}\left(\frac{2\pi amn}{2q+1}\right).
\eeq
Now, using the explicit expression for the Bessel function $K_{\frac32}(x)$, the corresponding contribution to the effective action reads
\beq\label{seffodd}
\delta S_{\mathrm{eff,}L(2q+1,1)}^{l=0} = \frac{1}{\pi}\sum_{r=1}^{q}r\sum_{n=1}^{\infty}\frac{1}{n^2}\left(\frac{2\pi amn}{2q+1}+1\right)\sin{\left(\frac{4\pi rn}{2q+1}\right)}e^{-\frac{2\pi amn}{2q+1}}.
\eeq

\subsection{$L(2,1)$}

For the simplest non-trivial case  $L(2,1)=S^3/\mathbb{Z}_2$, the degeneracies are nothing but the odd-$k$ degeneracies on the sphere, and the zeta function reads
\beq\nn
\zeta_{L(2,1)}(s) &=& \mu^{2s} \sum_{l=-\infty}^{\infty} \sum_{\substack{k=1\\k \mathrm{\,odd}}}^{\infty} k^2 \left[\left(\frac{k}{a}\right)^2 + m^2 + \left(\frac{2\pi l}{\beta}\right)^2 \right]^{-s} \\ \nn
&=& \mu^{2s} \sum_{l=-\infty}^{\infty} \left\{ \,\sum_{k=1}^{\infty} k^2 \left[\left(\frac{k}{a}\right)^2 + m^2 + \left(\frac{2\pi l}{\beta}\right)^2\right]^{-s}  \right. \\ \nn
& & -  \sum_{\substack{k=1\\k \mathrm{\,even}}}^{\infty}\left.k^2 \left[\left(\frac{k}{a}\right)^2 + m^2 + \left(\frac{2\pi l}{\beta}\right)^2\right]^{-s}\, \right\} \\
& = & \zeta_{S^3}(s;\beta,a,m) -\mu^{2s}\sum_{\substack{l=-\infty\\k=1}}^{\infty} (2k)^2 \left[ \left(\frac{2k}{a}\right)^2 + m^2 + \left(\frac{2\pi l}{\beta}\right)^2 \right]^{-s} \\ \nn
& = & \zeta_{S^3}(s;\beta,a,m) - 2^{2-2s}\zeta_{S^3}\left(s;2\beta,a,\frac{m}{2}\right)\,,
\eeq
i.e.
\beq\label{sefftwo}
S_{\mathrm{eff},L(2,1)}(\beta,a,m) = S_{\mathrm{eff},S^3}(\beta,a,m) - 4S_{\mathrm{eff},S^3}\left(2\beta,a,\frac{m}{2}\right)\,.
\eeq
We can directly obtain the effective action in the different limits from the corresponding expressions on the sphere. In particular, the subtractions needed to renormalize the theory at zero temperature can be read off from the ones already determined for $S^3$.

\subsection{Even-order lens spaces $L(2q,1)$}

To compute the $\beta$-independent corrections to the effective action on the even-order lens space
$L(2q,1)$ we rewrite the degeneracies of the eigenvalues $(2k+1)^2$ of the conformal Laplacian as
\beq
d_{2k+1} = \frac{(2k+1)^2}{q} + \frac{q-r-1}{q}(2k+1) - \frac{r}{q}(2k+1),
\eeq
for any $k=nq+r$ with $r=0,1, \dots, q-1$.
The zeta function becomes
\beq\nn
\zeta_{L(2q,1)}(s) &=& \mu^{2s}\sum_{\substack{k=0\\l=-\infty}}^{\infty} \frac{(2k+1)^2}{q}\left[\left(\frac{2k+1}{q}\right)^2 + m^2 + \left(\frac{2\pi l}{\beta}\right)^2\right]^{-s} + \delta\zeta_{L(2q,1)}(s) \\ \label{zetaeven}
& = & \frac{1}{q}\zeta_{L(2,1)}(s) + \delta\zeta_{L(2q,1)}(s),
\eeq
with
\beq\nn
\delta\zeta_{L(2q,1)}(s) \!=\! \mu^{2s}\! \sum_{\substack{n=0\\l=-\infty}}^{\infty} \sum_{r=0}^{q-1} \left( \frac{q\!-\!r\!-\!1}{q}\! -\! \frac{r}{q} \right) (2nq\!+\!2r\!+\!1) \left[ \left(\frac{2nq\!+\!2r\!+\!1}{a}\right)^2\! +\! m^2 \!+\! \left(\frac{2\pi l}{\beta}\right)^2\right]^{-s}.
\eeq

Redefining $q-r-1\rightarrow r$ in the first term and extracting some overall factors, we obtain
\beq\nn
& & \delta\zeta_{L(2q,1)}(s)\! =\! \sum_{\substack{n=0\\l=-\infty}}^{\infty} \left\{ \sum_{r=0}^{q-1} \frac{r}{q} \left(n\!+\!1\!-\!\frac{2r\!+\!1}{2q}\right) \left[ \left(n\!+\!1\!-\!\frac{2r\!+\!1}{2q}\right)^2 \!+\! \left(\frac{am}{2q}\right)^2 \!+\! \left(\frac{2\pi a l}{2q\beta}\right)^2\right]^{-s} \right.  \\ \nn
& & \,\,\,\,\,- \left. \sum_{r=0}^{q-1} \frac{r}{q} \left(n+\frac{2r+1}{2q}\right) \left[ \left(n+\frac{2r+1}{2q}\right)^2 + \left(\frac{am}{2q}\right)^2 + \left(\frac{2\pi a l}{2q\beta}\right)^2\right]^{-s} \right\}  2q\left(\frac{\mu a}{2q}\right)^{2s} .
\eeq

As in the case of odd-order lens spaces, it is easy to check that the only subtractions needed to renormalize the theory are the ones on $L(2,1)$, modded out by the volume factor $q$. At high temperatures the $l\neq 0$ terms in $\delta\zeta_{L(2q,1)}$ contribute to the effective action as
\beq\nn
\exp\left(-\frac{2\pi a n}{2q}\sqrt{m^2+\left(\frac{2\pi l}{\beta}\right)^2}\right)\,,
\eeq
which is exponentially suppressed with the temperature. Thus, one can consider only the $l=0$ terms in $\delta\zeta_{L(2q,1)}$, which can be written as
\beq\nn
\delta\zeta_{L(2q,1)}^{l=0}(s) &=& -\frac{2q}{1-s}\left(\frac{\mu a }{2q}\right)^{2s} \sum_{n=-\infty}^{\infty}\sum_{r=1}^{q-1} r \left. \frac{d}{d\alpha}\right\vert_{\alpha=1} \left[ \left(n+\frac{2r+\alpha}{2q}\right)^2 + \left(\frac{am}{2q}\right)^2 \right]^{-s+1}
\eeq
or, using Schwinger's proper-time representation and a Poisson summation in the $n$-sum,
\beq\nn
\delta\zeta_{L(2q,1)}^{l=0}(s) &=&{ \frac{2\pi^{\frac12}q}{(s-1)\Gamma(s-1)}\left(\frac{\mu a}{2q}\right)^{2s} }\sum_{r=1}^{q-1} r \left. \frac{d}{d\alpha}\right\vert_{\alpha=1} \left\{ \left(\frac{am}{2q}\right)^{3-2s}\Gamma(s-\tfrac32) \right. + \\ \nn
& & \left.+  {  4\left(\frac{am}{2\pi q}\right)^{\frac32-s}}\sum_{n=1}^{\infty} n^{s-\frac32} \cos{\left(2\pi\frac{2r+\alpha}{2q}n\right)}K_{s-\frac32}\left(\frac{2\pi am n}{2q}\right) \right\}\\ \nn
&=& - {  \frac{8s}{\Gamma(s\!+\!1)}\left(\!\frac{\pi\mu^2 a}{2qm}\!\right)^{\!s} \left(\!\frac{am}{2q}\!\right)^{\!\frac32}}\sum_{r=1}^{q-1} r \sum_{n=1}^{\infty}  n^{s-\frac12} \sin{\left(\!2\pi \frac{2r\!+\!1}{2q}n\!\right)} K_{s-\frac32}\left(\!\frac{2\pi am n}{2q}\!\right).
\eeq

Performing the $s$-derivative and using the explicit expression for $K_{\frac32}(x)$ we obtain for the corresponding contribution to the effective action
\beq\label{seffeven}
\delta S_{\mathrm{eff,}L(2q,1)}^{l=0} = \frac{1}{\pi} \sum_{r=1}^{q-1} r \sum_{n=1}^{\infty} \frac{1}{n^2} \left(\frac{2\pi amn}{2q}+1\right) \sin{\left(2\pi\frac{2r+1}{2q}n\right)}e^{-\frac{2\pi amn}{2q} }.
\eeq

\subsection{Prism spaces}

Using the non-vanishing degeneracies in \eqref{prism}, we can write the zeta function on the prism space $S^3/D_{4p}^*$ as
\beq\nn
\zeta_{S^3/D_{4p}^*}(s) & = & \mu^{2s}\sum_{l=-\infty}^{\infty} \left\{ \sum_{\substack{k=0\\ k \mathrm{\,even}}}^{\infty} (2k+1)\left(\left[\frac{k}{p}\right]+1\right) + \sum_{\substack{k=1\\ k \mathrm{\,odd}}}^{\infty} (2k+1)\left[\frac{k}{p}\right] \right\}  \left(\lambda_{2k+1} + \omega_l^2 \right)^{-s} \\[2mm]\label{zetaprism}
& = & \frac{1}{2}\zeta_{L(2p,1)}(s)+\delta\zeta_{S^3/D_{4p}^*}(s)\,,
\eeq
with
\beq\nn
\!\!\! \delta\zeta_{S^3/D_{4p}^*}(s)\!\!& = &\!\! { \frac{\mu^{2s}}{2}}\sum_{l=-\infty}^{\infty} \left(\,\sum_{\substack{k=0\\ k \mathrm{\,even}}}^{\infty} - \sum_{\substack{k=1\\ k \mathrm{\,odd}}}^{\infty}\,\right) (2k+1)  \left(\lambda_{2k+1} + \omega_l^2 \right)^{-s}\\
& = &{   \frac{\mu^{2s}}{2}}\sum_{l=-\infty}^{\infty}\sum_{k=0}^{\infty} \left\{  { (4k\!+\!1)\left(\lambda_{4k+1}\!+\!\omega_l^2\right)^{-s} -} { (4k\!+\!3)\left(\lambda_{4k+3}\!+\!\omega_l^2\right)^{-s}}\right\}\,.
\eeq
Redefining $k\rightarrow -k-1$ in the second sum, this piece of the zeta function can be written as
\beq\label{dzetaprism}
\delta\zeta_{S^3/D_{4p}^*}(s) & = & \frac{\mu^{2s}}{2}\sum_{k,l=-\infty}^{\infty} (4k+1)\left(\lambda_{4k+1}+\omega_l^2\right)^{-s}\,.
\eeq

Again, the only required subtraction is that of  the sphere, with the adequate volume factor.
Also, at high temperature the $l\neq 0$ terms in \eqref{dzetaprism} correspond to terms in the effective action that are exponentially suppressed with the temperature. The remaining terms can be written as
\beq
\delta\zeta_{S^3/D_{4p}^*}^{l=0}(s) & = & \frac{\left(\mu a\right)^{2s}}{4(1-s)} \sum_{k=-\infty}^{\infty} \left.\frac{d}{d\alpha}\right\vert_{\alpha=1} \left[ \left(4k+\alpha\right)^2 + \left(am\right)^2 \right]^{-s+1}\,
\eeq
or, using Schwinger's proper-time representation and a Poisson summation in the $k$-sum,
\beq\nn
\delta\zeta_{S^3/D_{4p}^*}^{l=0}(s) & = & -\frac{\pi^{\frac12}\left(\mu a\right)^{2s}}{4\Gamma(s)} \sum_{k=1}^{\infty} \left.\frac{d}{d\alpha}\right\vert_{\alpha=1} \cos{\left(\frac{\pi}{2}\alpha k\right)} \left(\frac{4am}{\pi k}\right)^{\frac32-s} K_{\frac32-s}\left(\frac{\pi amk}{2}\right)\\
& = & \frac{\pi^{\frac32}\left(\mu a\right)^{2s}}{8\Gamma(s)}\sum_{\substack{k=1\\ k \mathrm{\,odd}}}^{\infty} (-1)^{\left[k/2\right]} \,k \left(\frac{4am}{\pi k}\right)^{\frac32-s} K_{\frac32-s}\left(\frac{\pi amk}{2}\right)\,.
\eeq
The $s$-derivative provides the corresponding contribution to the effective action,
\beq
\delta S_{\mathrm{eff,}S^3/D_{4p}^*}^{l=0}(s) = -\frac{1}{\pi}\sum_{\substack{k=1\\ k \mathrm{\,odd}}}^{\infty} \frac{(-1)^{\left[k/2\right]}}{k^2} \left(1+\frac{\pi a m k}{2}\right) e^{-\frac{\pi amk}{2}}\,.
\eeq

\subsection{Tetrahedral, octahedral and Poincar\'e dodecahedral spaces}

In the remaining cases, the zeta functions can be written using a cyclic decomposition \cite{Dowker041}, the effective actions resulting
\beq\label{seffpol}
S_{\mathrm{eff,}S^3/\Gamma} = \frac12 \sum_q S_{\mathrm{eff,}S^3/\mathbb{Z}_{2q}}-\frac12 S_{\mathrm{eff,}S^3/\mathbb{Z}_2}\,,
\eeq
where $2\pi/q$ are the rotation angles over the three symmetry axes in the corresponding $SO(3)$ subgroup. For our purposes, it is enough to recall that the values of $q$ are $(2,3,3)$ for $\Gamma^\ast=T^*$, $(2,3,4)$ for $\Gamma^\ast=O^*$ and $(2,3,5)$ for $\Gamma^\ast=I^*$.

\section{Holonomy entropy and Renormalization Group flow}

From the above calculations one  can  derive the entropy $S=(\beta \partial_\beta-1)S_{\rm eff}$ of the theory. At low temperature ($\frac{\beta}{a}\gg 1$) we obtain an exponentially small entropy
\beq\label{eltlm}
\!\!S\left(\Gamma^\ast,ma,{\frac{\beta}{a}}\right)\! =\! \sum_{k=1}^{\infty} d_k\left[\frac{\beta}{a} \sqrt{k^2\!+\!(am)^2}\frac{e^{-\frac{\beta}{a}\sqrt{k^2\!+\!(am)^2}}}{1\!-\!e^{-\frac{\beta}{a}\sqrt{k^2\!+\!(am)^2}}}\!-\!
\log\left(1\!-\!e^{-\frac{\beta}{a}\sqrt{k^2\!+\!(am)^2}}\right)\right]
\eeq
both in the large mass-volume and in the small mass-volume scenarios.

However, at high temperature $\frac{\beta}{ a}\ll 1$  the entropy picks up a temperature-independent subleading contribution. In the
case $\beta m< 2\pi$ the total entropy is given by
\beq\label{ehtlmo}
S\left(\mathbb{Z}_{2q+1},ma,{\frac{\beta}{a}}\right)  &=& \frac{\pi}{6 (2q+1)} (am)^3 +\frac{4\pi^4}{45(2q+1)}\left(\frac{a}{\beta}\right)^3 - \frac{\pi^2}{6(2q+1)} (am)^2 \frac{a}{\beta}\nonumber\\
&-& \frac{1}{4\pi^2 (2q+1)}\sum_{k=1}^{\infty} \frac{1}{k^3} \left[ 1 + 2\pi ma k + \frac12 (2\pi mak)^2 \right] e^{-2\pi ma k}\\
&-&\frac{1}{\pi}\sum_{r=1}^{q}r\sum_{n=1}^{\infty}\frac{1}{n^2}\left(\frac{2\pi amn}{2q+1}+1\right)\sin{\left(\frac{4\pi rn}{2q+1}\right)}e^{-\frac{2\pi amn}{2q+1}} +{\cal O}({\beta})\nn
\eeq
for odd-order lens spaces $L(2q+1,1)$, whereas for even-order lens spaces $L(2q,1)$ it reads
\beq\label{ehtlme}
S\left(\mathbb{Z}_{2q},ma,{\frac{\beta}{a}}\right)  &=& \frac{\pi}{6 (2q)} (am)^3 +\frac{4\pi^4}{45(2q)}\left(\frac{a}{\beta}\right)^3 - \frac{\pi^2}{6(2q)} (am)^2 \frac{a}{\beta}\nonumber\\
&-& \frac{1}{4\pi^2 q}\sum_{k=1}^{\infty} \frac{1}{k^3} \left[ 1 + 2\pi ma k + \frac12 (2\pi mak)^2 \right] e^{-2\pi ma k}\nn\\
&+& \frac{1}{\pi^2 q}\sum_{k=1}^{\infty} \frac{1}{k^3} \left[ 1 + \pi ma k + \frac12 (\pi mak)^2 \right] e^{-\pi ma k}\\
&-& \frac{1}{\pi} \sum_{r=1}^{q-1} r \sum_{n=1}^{\infty} \frac{1}{n^2} \left(\frac{2\pi amn}{2q}+1\right) \sin{\left(2\pi\frac{2r+1}{2q}n\right)}e^{-\frac{2\pi amn}{2q} }+{\cal O}({\beta})\,,\nn\eeq
and for prism spaces it is
\beq\label{ehtlmp}
S\left(D_{4p}^*,ma,{\frac{\beta}{a}}\right)  &=& \frac12 S\left(\mathbb{Z}_{2p},ma,{\frac{\beta}{a}}\right) +\frac{1}{\pi}\sum_{\substack{k=1\\ k \mathrm{\,odd}}}^{\infty} \frac{(-1)^{\left[k/2\right]}}{k^2} \left(1+\frac{\pi a m k}{2}\right) e^{-\frac{\pi amk}{2}}\,.\eeq

\begin{figure}[t]
\centering
  \includegraphics[width=10cm]{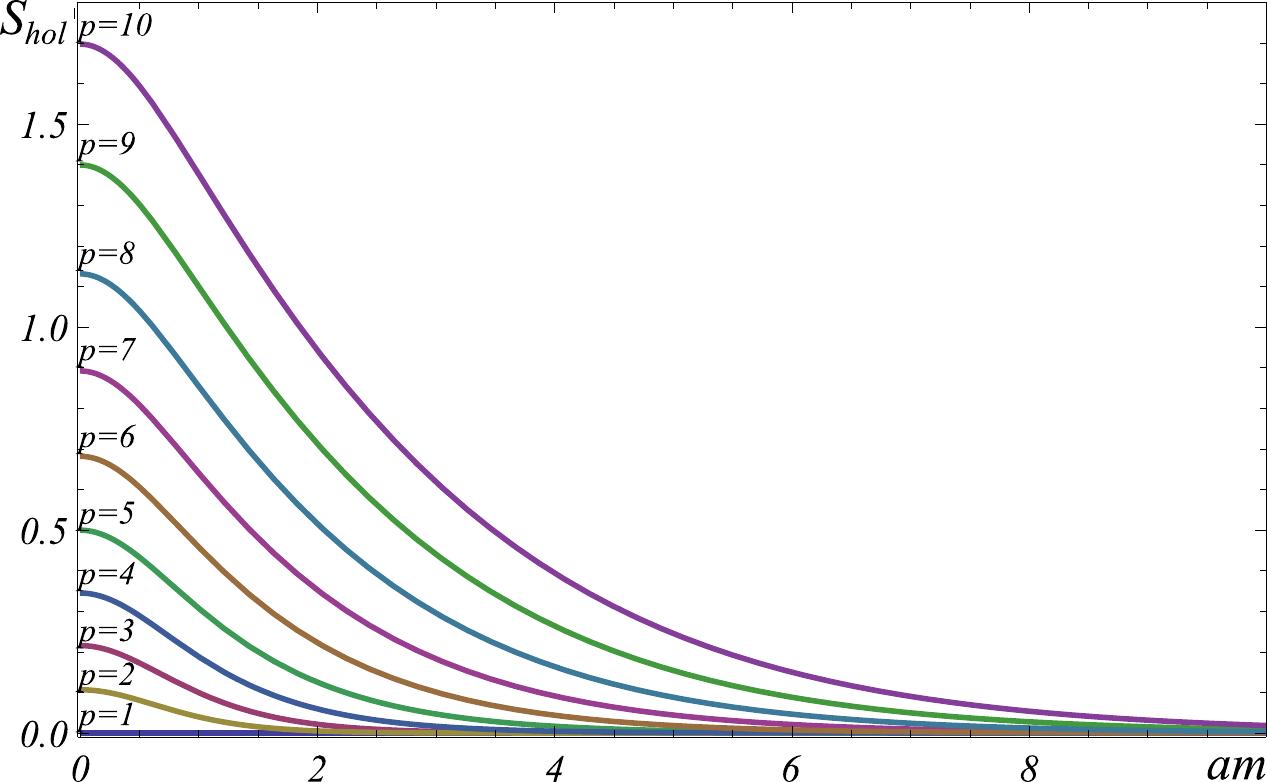}
  \caption{Behavior of the holonomy entropy $S_{\rm hol}{(p,ma)}$ with the mass for different lens spaces $L(p,1)$. In all cases $S_{\rm hol}$  is monotonically decreasing  with the mass, interpolating between the positive value of the holonomy entropy $S_{\rm hol}{(p,0)}$ at $m=0$ and its vanishing value at  $m=\infty$.  $S_{\rm hol}{(1,ma)}$ vanishes for all the mass values.} \label{massentrop}
\end{figure}

From the temperature-independent part of the entropy (which, except in the case of very large masses ($m\beta>2\pi$),
does coincide up to a sign with the Euclidean effective action of the same theory on the corresponding spherical space) we can extract the non-extensive part. We observe that, once this is done, there are  terms
proportional to $\frac1{p}$, some of which correspond to the contribution of a non-local density \cite{GZ}. In the quantum mechanical Euclidean path integral approach it can be shown that the remaining ones are due to contributions of non-contractible paths \cite{dW}. It is just for this reason that the contribution of this remainder defines the holonomy entropy $S_{\rm hol}$.
It can be directly obtained from the full thermodynamic entropy on the space by subtracting the full entropy of $S^3$ multiplied by the volume factor $1/|\Gamma^\ast_p|$ of the spherical space, i.e.

\beq\label{hehtlmo}
S_{\rm hol} (\Gamma^\ast_p,ma)= \lim_{\beta \to 0}\left[ S \left(\Gamma^\ast_p,ma,{\frac{\beta}{a}}\right) - \frac{1}{|\Gamma^\ast_p|} S \left(1,ma,{\frac{\beta}{a}}\right) \right]\,.
\eeq

\begin{figure}[t]
\centerline{\includegraphics[width=11cm]{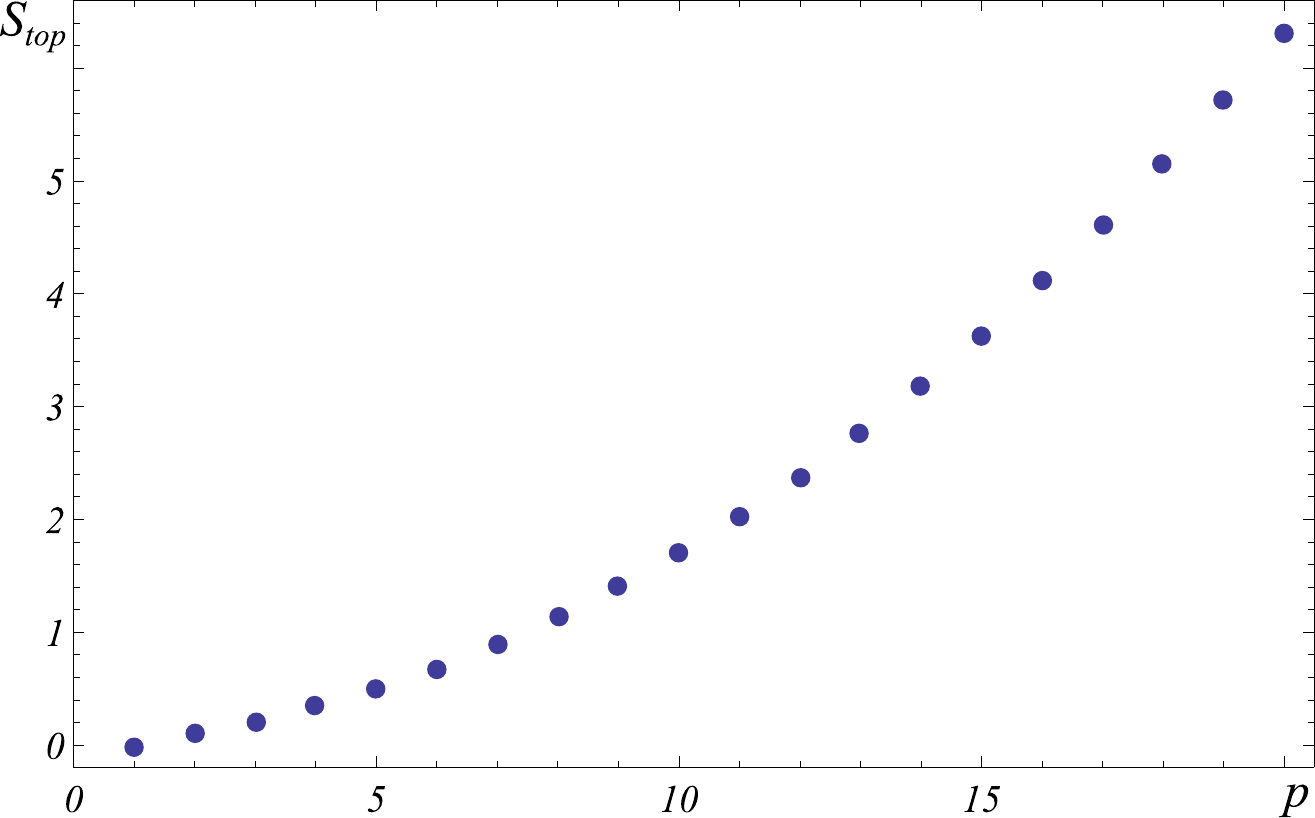}}
\caption{Dependence of the topological entropy $S_{\rm top}(p)=S_{\rm hol}(p,0)$   on the lens space $L(p,1)$. The dependence on $p$ is monotonically increasing with an asymptotic quadratic behavior for large $p$.}  \label{asymptrop}
\end{figure}

The holonomy entropy might depend on the metric of the space-time, but it is always finite.
If the mass of the scalar field is lower than the temperature, $m< \frac{2\pi}{\beta}$,
\beq\label{hehtsmo}
S_{\rm hol} (\mathbb{Z}_{2q+1},ma) &=&-\frac{1}{\pi}\sum_{r=1}^{q}r\sum_{n=1}^{\infty}\frac{1}{n^2}\left(\frac{2\pi amn}{2q+1}+1\right)\sin{\left(\frac{4\pi rn}{2q+1}\right)}e^{-\frac{2\pi amn}{2q+1}}\eeq
for odd-order lens spaces $L(2q+1,1)$, and by
\beq\label{hehtlme}
S_{\rm hol} (\mathbb{Z}_{2q},ma) &=& - \frac{1}{8\pi^2 q}\sum_{k=1}^{\infty} \frac{1}{k^3} \left[ 1 + 2\pi ma k + \frac12 (2\pi mak)^2 \right] e^{-2\pi ma k}\nn\\
&+ & \frac{1}{\pi^2 q}\sum_{k=1}^{\infty} \frac{1}{k^3} \left[ 1 + \pi ma k + \frac12 (\pi mak)^2 \right] e^{-\pi ma k}\nonumber \\
&-& \frac{1}{\pi} \sum_{r=1}^{q-1} r \sum_{n=1}^{\infty} \frac{1}{n^2} \left(\frac{2\pi amn}{2q}+1\right) \sin{\left(2\pi\frac{2r+1}{2q}n\right)}e^{-\frac{2\pi amn}{2q} }
\eeq
for even-order lens spaces $L(2q,1)$.

\begin{table}[t]
\begin{center}
\begin{tabular}{!{\vrule width 1.2pt}c|c|c|c|c|c|c!{\vrule width 1.2pt}}
\noalign{\hrule height 1.2pt}
p&1&2&3&4&5&6\\
\hline
$F$&0.030&-0.091&-0.205&-0.337&-0.494&-0.677\\
\hline
$S_{\rm top}$&0&0.107&0.215&0.345&0.500&0.682\\
\noalign{\hrule height 1.2pt}
\end{tabular}
\end{center}
\caption{Free energy and topological entropy for 4-d conformal scalars on lens spaces $L(p,1)$.}\label{table:estop4d}
\end{table}

\begin{figure}[b]
\centering
  \centerline{\includegraphics[width=11cm]{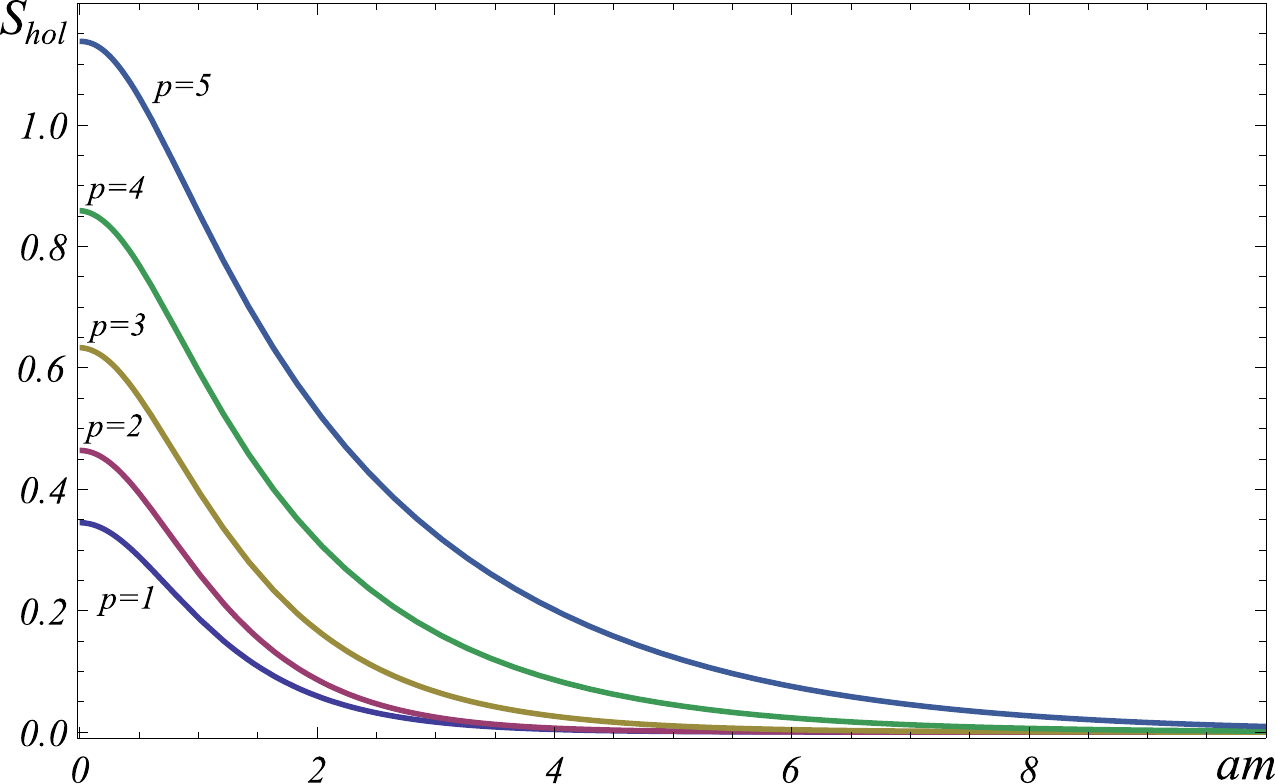}}
  \caption{Behavior of the holonomy entropy $S_{\rm hol}$ with the mass for the prism spaces $S^3/D_{4p}^*$. In all cases $S_{\rm hol}$  is monotonically decreasing  with the mass, interpolating between the positive value of the topological entropy $S_{\rm top}$ at $m=0$ and its vanishing value as  $m\rightarrow\infty$.} \label{sholprism}
\end{figure}

However, for large masses  $m>\frac{2\pi}{\beta}$, the holonomy entropy vanishes due to the decoupling of the heavy modes, which agrees with the
analytic continuation of \eref{hehtsmo} and \eref{hehtlme}. Now, if we look at the global behavior of
this entropy with $ma$, we observe that  the holonomy entropy  is a non-increasing function of $am$ for any lens space $L(p,1)$ (see figure \ref{massentrop}). Moreover, it vanishes exponentially as $ma\rightarrow \infty$. A very special case is the sphere $L(1,1)=S^3$, where the holonomy entropy vanishes for all the values of the mass.
At the $m=0$ conformal point, the holonomy entropy does coincide with the corresponding difference of topological entropies. We will denote such difference by $S_{\rm top}$,
\beq\label{tehtlmo}
S_{\rm top}(2q+1)=S_{\rm hol} (\mathbb{Z}_{2q+1},0)&=& -\frac{1}{\pi}\sum_{r=1}^{q}r\sum_{n=1}^{\infty}\frac{1}{n^2}\sin{\left(\frac{4\pi rn}{2q+1}\right)}\eeq
for $p=2q+1$ odd, and
\beq\label{tehtlme}
S_{\rm top}(2q)=S_{\rm hol} (\mathbb{Z}_{2q},0)=
 \frac78\frac{\zeta(3)}{\pi^2 q}
- \frac{1}{\pi} \sum_{r=1}^{q-1} r \sum_{n=1}^{\infty} \frac{1}{n^2} \sin{\left(2\pi\frac{2r+1}{2q}n\right)}
\eeq
for $p=2q$ even. In both cases, the topological entropy (holonomy entropy at the $m=0$ conformal point) is positive. The  values of this entropy for the first few lens spaces $L(p,1)$ are given in table \ref{table:estop4d} and displayed in figure \ref{asymptrop}. The positivity of $S_{\rm top}$ contrasts with the negative values of the  $F$ term of  free energies for
lens spaces with $p>1$, that are also listed in table \ref{table:estop4d}.
The fact that $F$ is positive only in the case $p=1$, i.e., only in the case involved in the $F$-theorem is quite remarkable.
The same behavior arises in the topological contribution to the vacuum energy $E_{\rm top}$: it is positive for $S^3$ and negative for all other spherical spaces  \cite{acm,Dowker89,Dowker041}.

When the mass of the field is the largest parameter of the theory, \eref{eltlm} implies that the topological entropy vanishes.

On the prism space $S^3/D_{4p}^*$, for  $m>\frac{2\pi}{\beta}$, the holonomy entropy is given by
\beq\label{hehtlmp}
S_{\rm hol} (D_{4p}^*,ma) &=& \frac12 S_{\rm hol} (\mathbb{Z}_{2p},ma) +\frac{1}{\pi}\sum_{\substack{k=1\\ k \mathrm{\,odd}}}^{\infty} \frac{(-1)^{\left[k/2\right]}}{k^2} \left(1+\frac{\pi a m k}{2}\right) e^{-\frac{\pi amk}{2}}\,,
\eeq
and, at the $m=0$ conformal point, it also coincides with the corresponding topological entropy,
\beq
S_{\rm hol} (D_{4p}^*,0) &=&  \frac{G}{\pi} + \frac{7}{16}\frac{\zeta(3)}{\pi^2 p}
- \frac{1}{2\pi} \sum_{r=1}^{p-1} r \sum_{n=1}^{\infty} \frac{1}{n^2} \sin{\left(2\pi\frac{2r+1}{2p}n\right)}\,,
\eeq
$G$ being Catalan's constant. Figure \ref{sholprism} shows the monotonically decreasing behavior of $S_{\rm{hol}}$ with $ma$ for some values of $p$.

\begin{figure}[t]
  \centerline{\includegraphics[width=11cm]{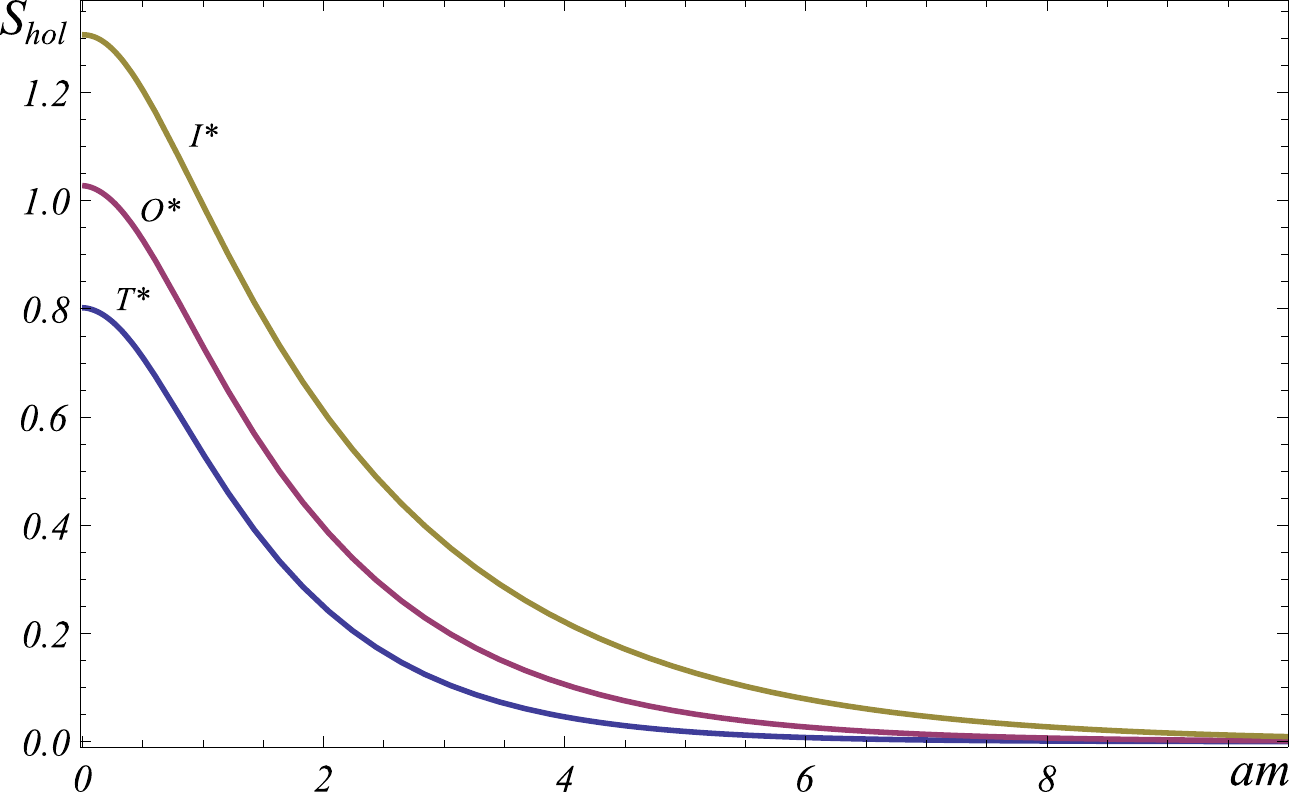}}
  \caption{Behavior of the holonomy entropy $S_{\rm hol}$ with the mass for octahedral $S^3/T^*$, truncated cube $S^3/O^*$ and Poincar\'e dodecahedral $S^3/I^*$ spaces. In the three cases
 $S_{\rm hol}$  is monotonically decreasing  with the mass, interpolating between the positive value of the topological entropy $S_{\rm top}$ at $m=0$ and its vanishing value at  $m=\infty$.} \label{2b}
\end{figure}

The holonomy entropy for the remaining spherical spaces can be obtained from \eqref{seffpol}. Its behavior with $ma$ is displayed in figure \ref{2b}. We remark that, again, $S_{\rm{hol}}$ is monotonically decreasing with $ma$.

\medskip

Because of the absence of interactions the behavior of the theory under the RG flow is Gaussian, and  given by the flow of an effective mass $m(\mu)$ running with the renormalization scale $\mu$ as \cite{wilson}
$$m(\mu)= m\frac{\mu_0}{\mu},$$
$\mu_0$ being the renormalization scale where the mass of the system is fixed $m(\mu_0)=m$ by the renormalization scheme.

Thus, the Wilsonian RG defined by the running of the holonomy entropy with the RG scale shows that $S_{\rm hol}(\mu)$ flows from  the positive value of the topological entropy at the UV ($\mu\gg\mu_0$) towards a zero value in the IR ($\mu\ll\mu_0$), i.e., it follows the expected behavior for a c-theorem,
\beq \label{ineq} S_{\rm top}^{UV}>S_{\rm top}^{IR}.\eeq
Moreover, the RG flow generates an irreversible flow of the holonomy entropy at intermediate scales
\beq \label{irr}
 \mu\frac d{d\mu} S_{\rm hol} (p,m(\mu)a)\geq 0\,.
\eeq

\section{Massless Dirac fermions on spherical backgrounds}

One important test of the conjecture is to check whether it holds or not for other field theories. In this section, we analyze the case of massless Dirac fermions.

To compute the topological entropy we only need the eigenvalues of the Dirac operator $\Dsl$ for four-component spinors on the spherical space $S^3/\Gamma^\ast$, which are
\cite{baar,Dowker041,teh}
\beq
\lambda^\pm_k &=& \pm \frac1{a}\left(k+1/2\right), \qquad k=0,1,2,\ldots,\infty.
\label{eigenvaluesf}
\eeq
The degeneracies are given by
\beq
d^\pm_k=k(k+1)
\eeq
for spinors on the sphere, by
\beq\nn
d^\pm_{2k+1} &=& {(k+1)}\left(2\left[\frac{2k+1}{2 q}\right]+1\right), \quad k=0,1,2,\ldots,\infty\\ \nn
d^\pm_{2k} &=& {k}\left(2[\frac{2k}{2 q}]+1\right),  \qquad\  \qquad \, \ \  k=0,1,2,\ldots,\infty.
\label{degeneraciesff}
\eeq
on even-order lens spaces $L(2 q,1)$, and by
\begin{align*}
d_k^+= d_k^-=
\begin{cases}   k\left(2\left[\frac{k}{2q+1}\right]+1\right) +\left[\frac{k}{2q+1}\right] &  \quad \mbox{for $r$ even ($\neq 2q$) } \\
  \phantom{\Biggl[}\!\!\!\!2 k\left(\left[\frac{k}{2q+1}\right]+1\right) +\left[\frac{k}{2q+1}\right] &  \quad \mbox{for $r = 2q$ }\\
              k\left(2\left[\frac{k}{2q+1}\right]+1\right) +\left[\frac{k}{2q+1}\right]+1 & \quad \mbox{for $r$ odd,}\end{cases}
              \end{align*}
for odd-order lens spaces $L(2q+1,1)$, where $k= (2q+1)\left[\frac{k}{2q+1}\right]+r$  with $r=0,1,\ldots,2q$.

We first note that, because of the absence of spectral asymmetry for 4-component massless Dirac fermions,
 $$\det \Dsl=\sqrt{\det\Dsl^2}\,,$$
so we can ignore the negative signs of the eigenvalues $\lambda^-_k$.

The corresponding $\zeta$-function contributions are given by
\begin{align}\nn
  \zeta_{S^3/\mathbb{Z}_{2q}} (s) =& 4 (2q)^{1-s} \sum_{r=0}^{q-1} \left\{\zeta_H\left(s-2,\frac{2r+\tfrac32}{2q}\right) + \left(\frac12 - \frac{2r+1}{2q}\right) \zeta_H\left(s-1,\frac{2r+\tfrac32}{2q}\right) + \right. \\ \nn
 & +  \frac{1}{4q}\left(\frac12 - \frac{2r+\tfrac32}{2q}\right) \zeta_H\left(s,\frac{2r+\tfrac32}{2q}\right) +\zeta_H\left(s-2,\frac{2r+\tfrac12}{2q}\right) +  \\ \label{zetaeven4}
& + \left. \left(\frac12 \!-\! \frac{2r\!+\!1}{2q}\right) \zeta_H\left(s\!-\!1,\frac{2r\!+\!\tfrac12}{2q}\right) \!-\! \frac{1}{4q}\left(\frac12 \!-\! \frac{2r\!+\!\tfrac12}{2q}\right) \zeta_H\left(s,\frac{2r\!+\!\tfrac12}{2q}\right) \right\}
\end{align}
for the even-order lens space $L(2q,1)$, and by
\begin{align}\nn
\!\!\!  \zeta_{S^3/\mathbb{Z}_{2q+1}} (s) \!=\! & 4 (2q+1)^{1-s} \left\{ \sum_{r=0}^{q-1} \left[ \zeta_H\left(s\!-\!2,\frac{2r\!+\!\tfrac12}{2q\!+\!1}\right) + \left(\frac12 \!-\! \frac{2r\!+\!\tfrac12}{2q\!+\!1}\right) \zeta_H\left(s\!-\!1,\frac{2r\!+\!\tfrac12}{2q\!+\!1}\right)  \right. \right. \\ \nn
 & -  \frac{1}{4} \frac{1}{2q+1} \zeta_H\left(s,\frac{2r+\tfrac12}{2q+1}\right) +\zeta_H\left(s-2,\frac{2r+\tfrac32}{2q+1}\right) +  \\ \nn
& + \left. \left(\frac12 - \frac{2r+\tfrac32}{2q+1}\right) \zeta_H\left(s-1,\frac{2r+\tfrac32}{2q+1}\right) + \frac{1}{4} \frac{1}{2q+1} \zeta_H\left(s,\frac{2r+\tfrac32}{2q+1}\right) \right] + \\ \label{zetaodd4}
& \!+\! \left. \zeta_H\!\left(\!\!s\!-\!2,\frac{2q\!+\!\tfrac12}{2q\!+\!1}\right) \!-\!\frac{q}{2q\!+\!1} \zeta_H\!\left(\!s\!-\!1,\frac{2q\!+\!\tfrac12}{2q\!+\!1}\right) \!-\!  \frac{1}{4} \frac{1}{2q\!+\!1} \zeta_H\!\left(\!s,\frac{2q\!+\!\tfrac12}{2q\!+\!1}\right) \right\}
\end{align}
for the odd-order lens space $L(2q+1,1)$.
The numerical values of the corresponding contributions to the free energy and topological entropy for the lowest  order lens spaces are listed in table \ref{table:estopf4d}.
\begin{table}[t]
\begin{center}
\begin{tabular}{!{\vrule width 1.2pt}c|c|c|c|c|c|c!{\vrule width 1.2pt}}
\noalign{\hrule height 1.2pt}
p&1&2&3&4&5&6\\
\hline
$F$&0.438&0.219&-0.038&-0.408&-0.903&-1.524\\
\hline
$S_{\rm top}$&0&0&0.184&0.518&0.990&1.597\\
\noalign{\hrule height 1.2pt}
\end{tabular}
\end{center}
\caption{Free energy and topological entropy of massless Dirac fermions on lens spaces $L(p,1)$.}\label{table:estopf4d}
\end{table}%

Notice that all the values of the topological entropy are positive and slightly larger than those of complex conformal scalars, which are twice those of real conformal scalars contained in table \ref{table:estop4d}.
The positivity of the topological entropy follows from the subadditivity property of the thermal entropy. In fact, such positivity follows from the subadditivity of the whole entropy in the high temperature limit of the 4-dimensional approach.

In summary, the topological entropy can be defined as a physical quantity associated to the dimensionally
reduced field theory, obtained as the infinite temperature limit of the original 4-dimensional theory, as the above calculations explicitly show.

\bigskip

\section{Three-dimensional fields on spherical backgrounds}

In order to have a pure 3-dimensional approach to the conjecture, it will be interesting to analyze the behavior of the holonomy entropy for three-dimensional conformal field theories.
There is a significant difference, both in case of scalars and of fermions. In the first case, the conformally invariant scalar theories are different in three and four dimensions. In four dimensions, the conformal coupling to the curvature $R$ is $\xi=\frac16$, whereas in three dimensions it is $\xi=\frac18$. In the case of fermions, the spinors have a different number of components: in 4-d theories they have four components, while in 3-d, the spinors in one irreducible representation have two; as we will see, the main difference is that the  4-d  Dirac operator is chiral symmetric, which implies a
symmetry between positive and negative eigenvalues that is absent in the 3-d case.

\bigskip
\subsection{Conformal scalar fields}

In the scalar case, the conformally invariant action is
\beq
S = \frac12 \int d^3x \sqrt{g} \left\{ g^{\mu\nu}\partial_{\mu}\phi\, \partial_{\nu}\phi + \frac{1}{8} R\, \phi^2\right\}.
\eeq
The eigenvalues of the relevant operator $-\Delta+\frac34$ on spherical spaces are
\beq
\lambda_k = \left(\frac{k}{a}\right)^2-\frac14, \quad k=1,2,\ldots,\infty\,,
\label{eigenvaluesbc}
\eeq
and the degeneracies are the same as in the 4-d case, eqs. \eref{evenw}-\eref{prism}.
An analogous calculation based on $\zeta$-function regularization method yields
\beq
F(2q+1)&=&
\frac{1}{16(2 q+1)} \left(\log 4-\frac{3 \zeta_R(3)}{\pi ^2}\right)
+\frac12\sum_{r=1}^q
\frac{r}{2 q+1}\left[\log\,\sin\frac{\pi(2 r+\tfrac{1}{2})}{2 q+1} \right. \\ \nn
& &- \left. \log\, \sin\frac{\pi (2 r-\tfrac{1}{2})}{2 q+1}\right]
+\frac12\sum_{r=1}^q  2r\left[\zeta'_H\left(-1,\frac{2 r -\tfrac{1}{2}}{2 q+1}\right)-\zeta'_H\left(-1,1-\frac{2 r +\tfrac{1}{2}}{2 q+1}\right) \right.\\ \nonumber
& & \left.+\zeta'_H\left(-1,\frac{2 r +\tfrac{1}{2}}{2 q+1}\right)-\zeta'_H\left(-1,1-\frac{2 r -\tfrac{1}{2}}{2 q+1}\right)\right]\,,
 \nonumber
 \eeq

 \vspace{-.2cm}
\beq
S_{\mathrm{hol}}(2q+1)&=&-\frac12\sum_{r=1}^q\frac{r}{2 q+1}\left[\log\, \sin\frac{\pi(2 r+\tfrac{1}{2})}{2 q+1}-\log\, \sin\frac{\pi (2 r-\tfrac{1}{2})}{2 q+1}\right] \\ \nn
& & - \frac12\sum_{r=1}^q 2r \left[\zeta'_H\left(-1,\frac{2 r -\tfrac{1}{2}}{2 q+1}\right)
-\zeta'_H\left(-1,1-\frac{2 r +\tfrac{1}{2}}{2 q+1}\right)+\zeta'_H\left(-1,\frac{2 r +\tfrac{1}{2}}{2 q+1}\right) \right.\\ \nn
& & -\left.\zeta'_H\left(-1,1-\frac{2 r -\tfrac{1}{2}}{2 q+1}\right)\right]\,,
 \nonumber
\eeq
for odd-order lens spaces $L(2q+1,1)$, and
\beq
\!\!\!F(2q)&=&\frac{1}{16(2 q)} \left(\log 4-\frac{3 \zeta_R(3)}{\pi ^2}\right) -\frac1{q} \left[\zeta'_H\left(-1,\frac14\right)-\zeta'_H\left(-1,\frac34\right)\right]
 \\ \nn &&-\sum_{r=1}^{q-1}
 r\left\{\zeta'_H\left(-1,1-\frac{2 r +\tfrac{1}{2}}{2 q}\right)-\zeta'_H\left(-1,\frac{2 r +\tfrac{1}{2}}{2 q}\right)+\zeta'_H\left(-1,1-\frac{2 r +\tfrac{3}{2}}{2 q}\right)\right.\\ \nn
& & -\zeta'_H\left(-1,\frac{2 r +\tfrac{3}{2}}{2 q}\right)
-\frac1{4q}\left[ \zeta'_H\left(0,1\!-\!\frac{2 r +\tfrac{1}{2}}{2 q}\right)
 +\zeta'_H\left(0,\frac{2 r +\tfrac{1}{2}}{2 q}\right) \right.\\ \nn
 & & -\left.\left.\zeta'_H\left(0,1\!-\!\frac{2 r +\tfrac{3}{2}}{2 q}\right)-\zeta'\left(0,\frac{2 r +\tfrac{3}{2}}{2 q}\right)\right]\right\}\,,
 \nonumber
 \eeq
 \beq\nn
S_{\mathrm{hol}}(2q)&=&\frac1{q} \left[\zeta'_H\left(-1,\frac14\right)-\zeta'_H\left(-1,\frac34\right)\right]
+\sum_{r=1}^{q-1}
r\left\{\zeta'_H\left(-1,1-\frac{2 r +\tfrac{1}{2}}{2 q}\right) \right.\\ \nn
& & -\zeta'_H\left(-1,\frac{2 r +\tfrac{1}{2}}{2 q}\right)+\zeta'_H\left(-1,1-\frac{2 r +\tfrac{3}{2}}{2 q}\right)-\zeta'_H\left(-1,\frac{2 r +\tfrac{3}{2}}{2 q}\right)
\nonumber\\ &&
-\frac1{4q}\left[ \zeta'_H\left(0,1-\frac{2 r +\tfrac{1}{2}}{2 q}\right)  +\zeta'_H\left(0,\frac{2 r +\tfrac{1}{2}}{2 q}\right) \right.\\ \nn
& & \left.\left.-\zeta'_H\left(0,1-\frac{2 r +\tfrac{3}{2}}{2 q}\right)-\zeta'_H\left(0,\frac{2 r +\tfrac{3}{2}}{2 q}\right)\right]\right\}\,,
 \nonumber
\eeq
for even-order lens spaces $L(2q,1)$. Table \ref{table:estop3d} displays the numerical values for the first few lens spaces, showing the positivity of the topological entropy for all of them.
\begin{table}[t]
\begin{center}
\begin{tabular}{!{\vrule width 1.2pt}c|c|c|c|c|c|c!{\vrule width 1.2pt}}
\noalign{\hrule height 1.2pt}
p&1&2&3&4&5&6\\
\hline
$F$&0.064&--0.114&-0.256&-0.407&-0.577&-0.772\\
\hline
$S_{\rm top}$&0&0.146&0.277&0.423&0.590&0.783\\
\noalign{\hrule height 1.2pt}
\end{tabular}
\end{center}
\caption{Free energy and topological entropy for 3-d conformal scalars in lens spaces $L(p,1)$ }\label{table:estop3d}
\end{table}%

Notice that the values of the topological entropy for 3-d conformal scalars are slightly larger than those of 4-d conformal scalars given in table \ref{table:estop4d}.
As before, the positivity follows from the subadditivity property of the topological entropy. In fact, it also follows from the subadditivity of the whole entropy in the high temperature limit of the 4-dimensional approach, even if in this case the theory from a 4-dimensional viewpoint is not conformally invariant.

From that point of view the topological entropy defined by pure three-dimensional conformal scalars on
spherical spaces can also be used as an entropic function for the RG flow. In the case of lens spaces $L(p,1)$, the definition of the topological entropy, eq. \eref{hehtlmo},
\beq\nn
S_{\rm top}(p)= S_{\rm hol} (p,0)= \lim_{\beta \to 0}\left[S(p,0,\beta)-\frac{1}{p}S(1,0,\beta)\right],
\eeq
is related to the topological  R\'enyi entanglement entropy obtained, for instance, in \cite{klebanov1},
\beq\nn
S(q)=\frac{q F(1)-F(q)}{1-q}\,.\eeq

In our case, $q=\frac{1}{p}<1$. So, $S_{\rm top}=\frac{p-1}{p}S({p})$.

Note that the $q\rightarrow 1$ limit of this R\'enyi entropy was used to establish the connection between the entanglement entropy and the $F$-theorem \cite{cas,Liu:2012eea}.
In our approach, the positivity of $S_{\rm top}(p)$ is a consequence of the subadditivity of the 4-d thermal entropy. Moreover, its non-decreasing dependence on $p$ is a consequence of a well-known property of the R\'enyi entropy.

\bigskip
\subsection{Free massless fermion field}

We consider a massless Dirac fermion field on $S^3/\Gamma^\ast$. The eigenvalues are again given by \eqref{eigenvaluesf}, but their degeneracies in this two-dimensional representation differ from  those of the four-component Dirac spinors \cite{baar,Dowker041,teh}.
In this case there is an asymmetry between positive and negative eigenvalues, which leads to the appearance of a parity anomaly \cite{redlich}. The effective action in spherical spaces has an imaginary contribution coming from the spectral asymmetry, which can be expressed in terms of the $\eta$-function. This contribution vanishes for the spherical geometry $S^3$, but not for non-trivial spherical factors $S^3/\Gamma^\ast$. However, this imaginary contribution is ambiguous and dependent on the choice of regularization and, in fact, it is possible to choose a regularization where it vanishes \cite{kogan}. We shall use such a regularization, which will allow us to  evaluate the genuine topological entropy from the identity
\beq\nn
| \det \Dsl|=\sqrt{\det\Dsl^2}.
\eeq

In the case of lens spaces $L(p,1)$ the degeneracies of the eigenvalues of the Dirac operator $\Dsl$ acting on two-component spinors depend on the even or odd character of $p$. For even-order lens spaces $L(2q,1)$ the degeneracies of two-component Dirac eigenspinors are \cite{baar,Dowker041}
\begin{align*}
d_k^+ = (k+1)
\begin{cases} 0 &  \quad \mbox{for $k$ even} \\
              2\left[\frac{k}{2q}\right]+1 & \quad \mbox{for $k$ odd,}\end{cases}\\
d_k^- = k
\begin{cases} 0 &  \quad \mbox{for $k$ odd} \\
              2\left[\frac{k}{2q}\right]+1 & \quad \mbox{for $k$ even.}\end{cases}
\end{align*}

The zeta function corresponding to $|\det \Dsl|$ is half the zeta function for the free massless Dirac fermion field in 4-d, eq. \eqref{zetaeven4}.

For odd-order lens spaces $L(2q+1,1)$ the degeneracies are
\begin{align*}
d_k^+ = (k+1)
\begin{cases}\left[\frac{k}{2q+1}\right] &  \quad \mbox{for $r$ even} \\
              \left[\frac{k}{2q+1}\right]+1 & \quad \mbox{for $r$ odd,}\end{cases}\\
d_k^- = k
\begin{cases} \left[\frac{k+1}{2q+1}\right] &  \quad \mbox{for $\bar{r}$ even} \\
             \left[\frac{k+1}{2q+1}\right]+1 & \quad \mbox{for $\bar{r}$ odd,}\end{cases}
\end{align*}
where $k= (2q+1)\left[\frac{k}{2q+1}\right]+r$ and $k+1= (2q+1)\left[\frac{k+1}{2q+1}\right]+\bar{r}$ with $r,\bar{r}=0,1,\ldots,2q$.

The zeta function corresponding to $|\det \Dsl|$ is, again, half  the obtained in the four-dimensional case, eq. \eqref{zetaodd4}.

The corresponding contributions to the free energy and topological entropy on lens spaces are summarized in table \ref{table:estopf3d}.

\begin{table}[t]
\begin{center}
\begin{tabular}{!{\vrule width 1.2pt}c|c|c|c|c|c|c!{\vrule width 1.2pt}}
\noalign{\hrule height 1.2pt}
p&1&2&3&4&5&6\\
\hline
$F$&0.219&0.109&-0.019&-0.204&-0.452&-0.762\\
\hline
$S_{\rm hol}$&0&0&0.092&0.259&0.495&0.798\\
\noalign{\hrule height 1.2pt}
\end{tabular}
\end{center}
\caption{Free energy and topological entropy of 3-d massless Dirac fermions in lens spaces $L(p,1)$.}\label{table:estopf3d}
\end{table}%
The results confirm that the topological entropy is positive, which is again a consequence of
the subadditivity property of the total entropy. The values of the topological entropy are again larger than those corresponding to 3-d complex conformal scalars, which are twice those corresponding to real conformal scalars, given in table \ref{table:estop4d}.

Thus, also in this case,  the topological entropy is an entropic function  of the RG flow,
once the parity anomaly contribution has been removed.

\bigskip

\section{Conclusions}

We have analyzed the RG flow of the  holonomy entropy $S_{\rm hol}$ which arises in the high temperature expansion
 of some massive theories. The analysis shows that this quantity behaves like a c-function under the RG flow, i.e., it
is monotonically decreasing towards the IR. In the conformal limits $m=0$ or $m=\infty$ the holonomy entropy $S_{\rm hol} $ does coincide with the topological entropy $S_{\rm top}$  introduced in \cite{Asorey:2012vp}, providing an inequality \eref{ineq}
which is reminiscent of similar inequalities that appear in even-dimensional space-times as  consequences of $c$-theorem in 1+1 dimensions and $a$-theorem in 3+1 dimensions.

These results suggest that it is quite conceivable that there is another generalization of c-theorems to odd-dimensional spaces, a point already raised in \cite{sm2,klebanov,cas}.
In the three-dimensional case, the generalization of the holonomy entropy for generic backgrounds is straightforward.
If we consider the  field theory in more general  Euclidean  compact space-times, the  effective action always contains a part
which is not divergent and can be associated with the contributions of non-contractible  paths on $M_3$ in the Schwinger-DeWitt
path integral approach \cite{dW}. This contribution becomes topological {in those cases where the theory is conformal or} there is an extension  of the theory to a
 $3+1$-dimensional conformal theory. Our conjecture is that the associated topological values satisfy the inequality \eref{ineq} and the flow
exhibits a monotonic behavior like \eref{irr} under any relevant perturbation. In other terms, the holonomy entropy  $S_{\rm hol}$
 is strongly monotonic along the RG flow. In connection with this point, it would be very interesting to analyze the behaviour of the holonomy entropy in interacting theories, like the $O(N)$ model. We postpone this study for future research.

\begin{figure}[t]
\centering
  \centerline{\includegraphics[width=9cm]{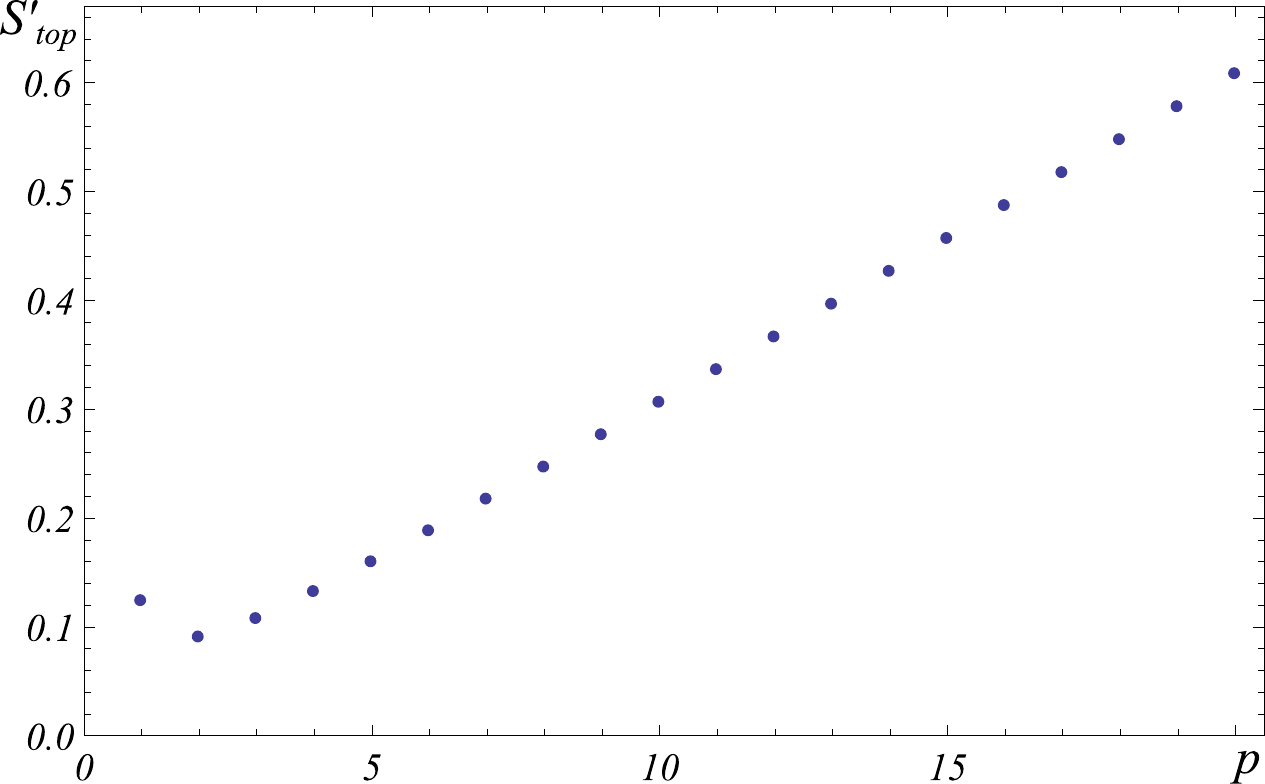}}
  \caption{\footnotesize{Dependence of the derivative of the topological entropy $S_{\rm top}(p)$ with respect to $p$ on the lens space $L(p,1)$. For $p\geq 2$, the dependence on $p$ is monotonically increasing, and the asymptotic behavior for large values of $p$ is linear.}} \label{deriv}
\end{figure}

Since there is no temperature dependence in the topological entropy, from a physical perspective, one might wonder whether or not the topological entropy has a  meaning as a physical entropy.
In the case of lens spaces it is possible to extend analytically the above results to non-integer values of $p$
by using the spectral function representation of the effective action for 4-dimensional conformal scalars \cite{Dowker12}
        \beq
S_{\rm top}^{(4)} (p) =
   \frac 1{4} \int_0^\infty  dx\  \Re \ \frac{p \sinh z + \cosh z \sinh pz}{z \sinh^2 z \sinh^2\frac{pz}{2}}
   - \frac 1{4 p} \int_0^\infty  dx\ \Re \ \frac{\sinh z\, ( 1+ \cosh z)}{z \sinh^2 z \sinh^2\frac{z}{2}},\label{continuous}\eeq
where the contour of integration $z=x+i\Delta$   is conveniently chosen in such a way that it   does not cross any singularity,
which occurs  for small enough values of   $\Delta$.

In that case, one
might speculate with the possible interpretation of $p$ as a kind of topological temperature. An explicit calculation shows that
the variation of the topological entropy with respect to this topological temperature does satisfy the second law of thermodynamics even
at very low values of $p$ (see figure \ref{deriv}). The meaning of this property is  unclear to us. However, the above approach opens new perspectives to the analysis of possible implications of the topological entropy in topological quantum computation with higher dimensional systems \cite{kitaev}.

 Moreover, there is an unexpected relation  between the $p$-derivative of the topological entropy
 of the 4-dimensional  conformal scalar field theory $\partial_p S^4_{\rm top}(p)$ at $p=1$ and the free energy $F$ of
 the 3-dimensional  conformal scalar field theory on $S^3$ involved in the $F$-theorem.  In appendix \ref{ap2} it is shown that
 \beq
 \partial_p S^{(4)}_{\rm top}(p)\Big|_{p=1} =2 F^{(3)}_{S^3}.
 \label{strange}
  \eeq
 This formula
 establishes a very interesting connection between conformal theories in $d=4$ and $d=3$ dimensions.
  The implications of this relation and its relationship with other similar connections
  between conformal theories in different dimensions \cite{gkl} will be further explored in a forthcoming paper.

\acknowledgments{We thank Ilya Shapiro for discussions.
Research of CGB, DD and EMS was partially supported by CONICET PIP681, ANPCyT PICT0605 and UNLP Proyecto 11/X615 (Argentina).
The work of MA and IC has been partially supported by  Spanish DGIID-DGA grant 2013-E24/2 and  Spanish
MICINN grants FPA2012-35453 and CPAN-CSD2007-00042.}

\appendix

\section{Casimir energy for the massive scalar on $S^3$}
\label{ap1}

Although the calculation of the Casimir energy for a massive scalar field on the $S^1\times S^3$ background has been performed earlier by many authors (see e.g. \cite{Ford,elizalde,myers} and references therein), we
include a derivation which emphasizes the matching of the standard value of the Casimir
energy for massless theories with {appropriate} renormalization prescriptions which are compatible with decoupling theorems.

The Casimir energy as a mode sum is given as the sum of the eigenvalues times its degeneracy,
\begin{equation}
E_C=\frac{1}{2}\sum_{k=1}^\infty k^2\sqrt{\left(\frac{k}{a}\right)^2+m^2}\,.
\end{equation}
Using the zeta-function regularization we can write
\begin{equation}
E_C(s)=\frac{1}{2}\mu^{2s+1}\sum_{k=1}^\infty k^2\left[\left(\frac{k}{a}\right)^2+m^2\right]^{-s}=\frac12\mu^{2s+1}a^{2s}\sum_{k=1}^\infty k^2\left[k^2+(am)^2\right]^{-s}\,,\label{cas-s}
\end{equation}
the Casimir energy being recovered for $s=-1/2$. The analytic extension of the sum \eqref{cas-s} is different for  large and  small volume-mass regimes.

\subsection{Large volume-mass case $ma\gg 1$}

We start from expression \eqref{cas-s} rewritten as
\beq
E_C(s)=\frac{1}{2}\mu^{2s+1}a^{2s}\sum_{k=1}^\infty \left\{\left[k^2+(am)^2\right]^{-s+1}-(am)^2\left[k^2+(am)^2\right]^{-s}\right\}\,.\label{cas-s2}
\eeq

The problem is reduced to the computation of a sum of the form
\beq
\sum_{k=1}^{\infty}\left[k^2+(am)^2\right]^{-r}\,,
\eeq
that can be achieved by using  Schwinger's proper-time representation and  Poisson-Jacobi inversion formula. The result is
\beq
\sum_{k=1}^{\infty}\left[k^2+(am)^2\right]^{-r} & = & \frac{1}{2} (am)^{-2r}+\frac{\sqrt{\pi}}{2}\frac{\Gamma(r-\tfrac12)}{\Gamma(r)}(am)^{-2r+1}\nn\\
& + & \frac{\sqrt{\pi}}{\Gamma(r)}\sum_{n=1}^\infty 2\,\left(\frac{n\pi}{am}\right)^{r-1/2} K_{r-\frac{1}{2}}(2n\pi m a)\,.\nonumber
\eeq

The regularized Casimir energy \eqref{cas-s} is
\beq
E_C(s) & = & \frac{1}{2}\mu^{2s+1}a^{2s}\left\{\frac{\sqrt{\pi}}{2}\frac{\Gamma(s-\tfrac32)}{\Gamma(s-1)}(am)^{-2s+3} +\frac{2\sqrt{\pi}}{\Gamma(s-1)}\sum_{n=1}^\infty \left(\frac{n\pi}{am}\right)^{s-\frac32}K_{s-\frac32}(2n\pi am)\right.\nonumber\\
& - &\frac{\Gamma(s-\tfrac12)}{\Gamma(s)}\frac{\sqrt{\pi}}{2}(am)^{-2s+3}- (am)^2\frac{\sqrt{\pi}}{\Gamma(s)}\left.\sum_{n=1}^\infty 2\left(\frac{n\pi}{am}\right)^{s-1/2} K_{s-\frac12}(2n\pi am)\right\}\,.
\eeq
The Casimir energy can be obtained letting  $s=-\tfrac12+\epsilon$,
\beq
E_C(\epsilon) & = & \left(\frac{\mu}{m}\right)^{2\epsilon}\frac{1}{2a}\left\{\frac{\sqrt{\pi}}{2}(am)^4
\left[\frac{\Gamma(\epsilon-2)}{\Gamma(\epsilon-\frac32)}-\frac{\Gamma(\epsilon-1)}{\Gamma(\epsilon-\frac12)}\right]\right.\nonumber\\
& + & \left.\frac{2\pi^{\epsilon-2+\tfrac12}}{\Gamma(\epsilon-\tfrac32)}(am)^{\epsilon+2}\sum_{n=1}^\infty \frac{K_{2-\epsilon}(2n\pi am)}{n^{2-\epsilon}}- \frac{2\pi^{\epsilon-\tfrac12}}{\Gamma(\epsilon-\tfrac12)}(am)^{\epsilon+3}\sum_{n=1}^\infty \frac{K_{1-\epsilon}(2n\pi am)}{n^{1-\epsilon}}\right\}\,,\nn
\eeq
and performing a  small-$\epsilon$ expansion,
\beq
E_C & = & \frac{(am)^4}{32a}\left(-\frac{1}{\epsilon}+2\log{\frac{m}{2\mu}}+\frac12\right)\nonumber\\
& + & \frac{3}{4\pi^2 a}(am)^2\sum_{n=1}^\infty \frac1{n^{2}}  K_2(2n\pi am)+\frac{1}{2\pi a}(am)^3\sum_{n=1}^\infty \frac1{n} K_1(2n\pi am)\,.\label{casimirbare}
\eeq

The appearance of a pole term requires a renormalization prescription based on a physical requirement that fixes the finite part. We remark that in the infinite volume $a\rightarrow \infty$ limit (which corresponds to flat Euclidean space) the contribution to the Casimir energy, given by the first line of \eqref{casimirbare}, does not vanish for any value of $\epsilon$ and $\mu$. Thus, a natural choice of the renormalization prescription should remove these terms. This renormalization prescription can be implemented by adding a counterterm to the bare Hamiltonian of the form
\beq\label{counterh}
\Delta H=-\frac{(am)^4}{32a}\left(-\frac{1}{\epsilon}+2\log\frac{m}{2\mu}+{\frac12}\right)\,.
\eeq
This renormalization condition agrees with the  normal ordering prescription in flat space. The final expression for the Casimir energy,
\beq
E_C=\frac{3}{4\pi^2 a}(am)^2\sum_{n=1}^\infty \frac1{n^{2}}  K_2(2n\pi am)+\frac{1}{2\pi a}(am)^3\sum_{n=1}^\infty \frac1{n} K_1(2n\pi am)\,,\label{casimirren}
\eeq
agrees with \eref{ce} in the low temperature regime for large volumes and heavy scalar fields.

\subsection{Small volume-mass case $ma\ll 1$}

In this case a binomial expansion of the sum \eref{cas-s} yields
\beq
E_C(s) = \frac{\mu}{2}(\mu a)^{2s} \sum_{n=0}^\infty \frac{\Gamma(-s+1)}{n!\Gamma(-s-n+1)} (am)^{2n} \zeta_R(2n+2s-2)\,.
\eeq
For $s\rightarrow \frac12$ the pole of the Riemann zeta function $\zeta_R$ in the last expression occurs when $n=2$.
Letting $s=-\tfrac12+\epsilon$, we have
\beq
E_C(\epsilon) & = & \frac{1}{2a}(\mu a)^{2\epsilon} \Bigg{\{} \zeta_R(2\epsilon -3) +\frac{\Gamma(-\epsilon+\tfrac32)}{\Gamma(-\epsilon+\tfrac12)}(am)^2\zeta_R(2\epsilon-1)
+\frac{\Gamma(-\epsilon+\tfrac32)}{2\Gamma(-\epsilon-\tfrac12)}(am)^4\zeta_R(2\epsilon+1)\nn\\
& + & \sum_{n=3}^\infty \frac{\Gamma(-\epsilon+\tfrac32)}{n!\Gamma(-\epsilon-n+\tfrac32)}(am)^{2n}\zeta_R(2\epsilon +2n-3)\Bigg{\}}\,.
\eeq
Finally, we expand around $\epsilon=0$ and add the same counterterm as in the large-$am$ case, obtaining for the Casimir energy
\beq
E_C & = & \frac{1}{240 a} -\frac{(am)^2}{48 a} -\frac{(am)^2}{16 a} \left(\log{\frac{am}{2}}+\gamma +\frac14 \right)\nn\\
 & + & \frac{1}{2a}\sum_{n=3}^{\infty} \frac{\Gamma(\tfrac32)}{n!\Gamma(\tfrac32 -n)}(am)^{2n}\zeta_R(2n-3)\,,
\eeq
which  agrees with the value obtained in the low temperature regime for small volumes and light mass scalar fields  \eref{seff2} (see also \cite{elizalde,myers}).
The fact that the renormalization prescription for the Casimir energy is the same in both extreme asymptotic mass regimes is surprising and can be understood thanks to the unification provided by the finite temperature approach. The introduction of an extra dimension provides an unique way of unifying all renormalization
prescriptions by adding a single counterterm to the effective action, which is independent of the asymptotic regime of $ma$.  The resulting renormalized effective action is finite for all values of $ma$ interpolating between the
different asymptotic regimes.
In particular, the same prescription holds in the high temperature regime for the calculation of
the topological entropy.

\section{Relation between  topological entropies of conformal scalar fields in $d=4$ and $d=3$}
\label{ap2}

The derivative with respect to $p$ of the topological entropy of 4-dimensional conformal scalars can be easily derived from
 formula \eref{continuous}, which provides an analytic extension of the topological entropy for continuous values of $p$
   \cite{Dowker12}. The result of this
 derivative  at $p=1$ is
\beq
 \left.\partial_p S^{(4)}_{\rm top}(p)\right\vert_{p=1} &=& \frac 1{4} \int_0^\infty  dx\  \Re \frac{\sinh z}{z \sinh^2 z \sinh^2\frac {z}{2}} + \frac 1{4} \int_0^\infty  dx\ \Re \ \frac{\sinh z\, ( 1+ \cosh z)}{z \sinh^2 z \sinh^2\frac{z}{2}}  \nn \\ \nn
&+&\frac 1{4} \int_0^\infty  dx\ \Re \frac{\cosh^2 z- \coth\frac{z}{2}\sinh z\, (1+ \cosh z)}{\sinh^2 z \sinh^2\frac {z}{2}} = I_1 + I_2 + I_3\,.\eeq

Let us prove the surprising result  \eref{strange}, which tells us that this quantity is twice  the free energy of the 3-d conformal scalar,
\beq\label{efe}
F^{(3)} = -\frac 1{4} \int_0^\infty  dx\ \Re \frac{\cosh\frac{z}{2}\,(1+\cosh z)}{z\sinh^2z \sinh^2\frac{z}{2}}\,,
\eeq
where now $z=x+i \Delta$  with $0<\Delta<2\pi$.

In order to prove the result we first remark that the third integral, $I_3$,  is  in fact vanishing.
The integral $I_3$ can be written as
\beq\nn
I_3 = -\frac 1{4} \int_0^\infty  dx\ \Re \frac{1+ 2\cosh z}{\sinh^2 z \sinh^2\frac {z}{2}}\,,
\eeq
since
\beq\nn
\cosh^2z-\coth\frac{z}{2}\sinh z \,(1+\cosh z) = -1-2\cosh z.
\eeq
The value of the integral is independent of $\Delta$ provided that  $0<\Delta<\pi$. Choosing $\Delta= \pi/2$, the integral becomes
\beq\nn
I_3 = -\frac 1{4} \int_0^\infty  dx\ \Re \frac{1+ 2\cosh \left(x + i\frac{\pi}{2}\right)}{\sinh^2 \left(x + i\frac{\pi}{2}\right) \sinh^2\left(\frac{x}{2} + i\frac{\pi}{4}\right)},
\eeq
and because of the identities
\beq\nn
\Re \frac{1+ 2\cosh \left(x + i\frac{\pi}{2}\right)}{\sinh^2 \left(x + i\frac{\pi}{2}\right) \sinh^2\left(\frac{x}{2} + i\frac{\pi}{4}\right)} = - \Re \frac{2+4i\sinh x}{\cosh^2 x (\sinh\frac{x}{2}+i\cosh\frac{x}{2})^2} = 2\left(\frac{3}{\cosh^4x} -\frac{2}{\cosh^2x}\right),
\eeq
and
\beq\nn
\int_0^\infty \frac{dx}{\cosh^2x} = 1\,,\,\,\,\, \int_0^\infty \frac{dx}{\cosh^4x} = \frac{2}{3}\,,
\eeq
we get that  finally  $I_3=0$.

Thus,
\beq\nn
\left.\partial_p S^{(4)}_{\rm top}(p)\right\vert_{p=1} = I_1+I_2 =  \frac 1{4} \int_0^\infty  dx\  \Re \frac{2+\cosh z}{z \sinh^2 z \sinh^2\frac {z}{2}}\,.
\eeq
Here $\Delta$ can have any value $0<\Delta<\pi$, but if we  choose $\Delta=\pi/2$,
\beq\nn
\Re \ \frac{2+\cosh z}{z \sinh^2 z \sinh^2\frac {z}{2}} = 2 \Re \ \frac{\sinh x-2i}{(x+i\frac{\pi}{2})\cosh x (\sinh\frac{x}{2}+i\cosh\frac{x}{2})^2} = \frac{2\pi-6x\sinh x -\pi\sinh^2x}{(x^2+\frac{\pi^2}{4})\cosh^3x}\,,
\eeq
and the integral becomes
\beq\nn
\left.\partial_p S^{(4)}_{\rm top}(p)\right\vert_{p=1} = \frac 1{4} \int_0^\infty  dx\  \frac{2\pi-6x\sinh x -\pi\sinh^2x}{(x^2+\frac{\pi^2}{4})\cosh^3x}\,.
\eeq

On the other hand, if we choose  $\Delta = \pi$ as the contour of integration in the expression  \eqref{efe} of 3-d conformal scalar free energy, the integral can be rewritten as
\beq\nn
F^{(3)} = -\frac 1{4} \int_0^\infty  dx\ \frac{\pi(1-\cosh x)}{2(x^2+\pi^2)\cosh^3\frac{x}{2}} = -\frac \pi{16} \int_0^\infty  dx\ \frac{1-\cosh 2x}{(x^2+\frac{\pi^2}{4})\cosh^3x},
\eeq
since,
\beq\nn
\Re \frac{\cosh\frac{z}{2}\,(1+\cosh z)}{z\sinh^2z \sinh^2\frac{z}{2}} = \Re \frac{i\sinh\frac{x}{2}\,(1-\cosh x)}{(x+i\pi)\sinh x \cosh^2\frac{x}{2}} = \frac{\pi\,(1-\cosh x)}{2(x^2+\pi^2)\cosh^3\frac{x}{2}}.
\eeq

\medskip

To prove the required identity \eref{strange} it is necessary to show  that the difference
\beq\nn
\left.\partial_p S^{(4)}_{\rm top}(p)\right\vert_{p=1}-2F^{(3)} = \frac{1}{8} \int_0^\infty  dx\ \frac{4\pi -12 x \sinh x -2\pi\sinh^2x+\pi-\pi\cosh 2x}{(x^2+\frac{\pi^2}{4})\cosh^3x}
\eeq
vanishes. Let us  rewrite the integral as
\beq\nn
 \frac{1}{2} \int_0^\infty \!\! dx\ \frac{4-2\cosh^2\frac{\pi x}{2}}{(x^2+1)\cosh^3\frac{\pi x}{2}} -\frac{1}{2} \int_0^\infty \!\! dx\ \frac{3 x \sinh\frac{\pi x}{2}}{(x^2+1)\cosh^3\frac{\pi x}{2}} = \tilde{I}_1 + \tilde{I}_2\,.
\eeq

The integrals $\tilde{I}_1$ and $\tilde{I}_2$ can be evaluated by means of the Laplace transforms
\beq\nn
\frac{1}{1+x^2} = \int_0^\infty dt\ e^{-t} \cos tx\,,\,\,\,\,\,\, \frac{x}{1+x^2} = \int_0^\infty dt\ e^{-t} \sin tx\,,
\eeq
which leads to
\beq\nn
\tilde{I}_1 = \int_0^\infty dt\ e^{-t} \int_0^\infty dx\ \frac{2-\cosh^2\frac{\pi x}{2}}{\cosh^3\frac{\pi x}{2}} \cos tx = \int_0^\infty dt\ e^{-t} \frac{4}{\pi^2}t^2 \,{\rm sech}\, t = \frac{3 \zeta_R(3)}{2\pi^2}\,,
\eeq
and
\beq\nn
\tilde{I}_2 = -\frac{3}{2}\int_0^\infty dt\ e^{-t} \int_0^\infty dx\ \frac{\sinh\frac{\pi x}{2}}{\cosh^3\frac{\pi x}{2}} \sin tx = -\frac{3}{\pi^2}\int_0^\infty dt\ e^{-t} t^2 \,{\rm csch}\, t = -\frac{3 \zeta_R(3)}{2\pi^2}\,,
\eeq
and concludes the proof of the surprising identity \eref{strange}.

\end{document}